\RequirePackage{rotating}
\documentclass[fleqn,usenatbib]{mnras}

\usepackage{newtxtext,newtxmath}

\usepackage[T1]{fontenc}
\usepackage{ae,aecompl}

\usepackage{graphicx}	
\usepackage[graphicx]{realboxes}
\usepackage{adjustbox}
\usepackage{subcaption}
\usepackage{pdflscape}

\usepackage{amsmath}	
\usepackage{rotating}

\title[GMCs in a simulated M51-like galaxy]{Simulations of the star-forming molecular gas in an interacting M51-like galaxy: cloud population statistics}

\author[R. G. Tress et al.]{Robin G. Tre{\ss},$^{1}$\thanks{E-mail: robin.tress@uni-heidelberg.de}
Mattia C. Sormani,$^{1}$
Rowan J. Smith,$^{2}$
Simon C. O. Glover,$^{1}$
\newauthor
Ralf S. Klessen,$^{1,3}$
Mordecai-Mark Mac Low,$^{4,5}$
Paul Clark,$^{6}$
and Ana Duarte-Cabral$^{6}$
\\
$^1$Universit\"{a}t Heidelberg, Zentrum f\"{u}r Astronomie, Institut f\"{u}r theoretische Astrophysik, Albert-Ueberle-Str. 2, 69120 Heidelberg, Germany \\
$^2$Jodrell Bank Centre for Astrophysics, Department of Physics and Astronomy, University of Manchester, Oxford Road, Manchester M13 9PL, UK\\
$^3$Universit\"at Heidelberg, Interdisziplin\"ares Zentrum f\"ur Wissenschaftliches Rechnen, Im Neuenheimer Feld 205, 69120 Heidelberg, Germany \\
$^4$Department of Astrophysics, American Museum of Natural History, 79th Street at Central Park West, New York, NY 10024, USA \\
$^5$Center for Computational Astrophysics, Flatiron Institute, 162 Fifth Avenue, New York, NY 10010, USA \\
$^6$School of Physics and Astronomy, Queen's Buildings, The Parade, Cardiff University, Cardiff, CF24 3AA, UK \\
}

\date{Accepted 2021 June 09. Received 2021 June 08; in original form 2020 December 14}

\pubyear{2021}

\begin{document}
\label{firstpage}
\pagerange{\pageref{firstpage}--\pageref{lastpage}}
\maketitle

\begin{abstract}
  To investigate how molecular clouds react to different environmental conditions at a galactic scale, we present a catalogue of giant molecular clouds resolved down to masses of $\sim 10$~M$_{\odot}$ from a simulation of the entire disc of an interacting M51-like galaxy and a comparable isolated galaxy. Our model includes time-dependent gas chemistry, sink particles for star formation and supernova feedback, meaning we are not reliant on star formation recipes based on threshold densities and can follow the physics of the cold molecular phase.  We extract giant molecular clouds from the simulations and analyse their properties. In the disc of our simulated galaxies, spiral arms seem to act merely as snowplows, gathering gas and clouds without dramatically affecting their properties. In the centre of the galaxy, on the other hand, environmental conditions lead to larger, more massive clouds. While the galaxy interaction has little effect on cloud masses and sizes, it does promote the formation of counter-rotating clouds. We find that the identified clouds seem to be largely gravitationally unbound at first glance, but a closer analysis of the hierarchical structure of the molecular interstellar medium shows that there is a large range of virial parameters with a smooth transition from unbound to mostly bound for the densest structures. The common observation that clouds appear to be virialised entities may therefore be due to CO bright emission highlighting a specific level in this hierarchical binding sequence. The small fraction of gravitationally bound structures found suggests that low galactic star formation efficiencies may be set by the process of cloud formation and initial collapse. 
\end{abstract}

\begin{keywords}
galaxies: ISM -- ISM: clouds -- ISM: structure -- hydrodynamics -- stars: formation -- ISM: kinematics and dynamics
\end{keywords}



\section{Introduction}
\label{sec:introduction}
Understanding the formation and dynamical evolution of the molecular phase in the interstellar medium (ISM) of galaxies is of fundamental importance for the study of star formation and galactic evolution, since it is within this phase that essentially all star formation occurs. Cooling in the molecular phase is efficient, and so it has low temperatures ($T<100$~K), and consequently a high density and low volume filling factor. A substantial fraction of the cold molecular phase is associated with giant molecular clouds (GMCs), as seen in CO observations of our own Galaxy and others. The dynamical state of the molecular phase is still far from fully understood and, in particular, a comprehensive picture of GMCs in a galactic context is missing. The nature of the dynamical processes that shape the ISM can be revealed by a statistical analysis of star-forming molecular gas; the study of GMC properties and their connection with the local galactic environment therefore remains an active research topic.

A major point of debate is centred on the dynamical state of GMCs. It is still discussed whether they are in (or close to) virial equilibrium, or whether they are freely collapsing gravitationally bound objects instead. If they are merely emergent structures in the ISM turbulent cascade, on the other hand, their properties would be set by the Mach number rather than a requirement of virial equilibrium \citep[e.g][]{Krumholz&McKee2005,    Hennebelle&Chabrier2011,Padoan&Nordlund2011, Federrath&Klessen2012,Burkhart2018}. Furthermore, the importance of environmental conditions for their dynamical state remains a central question. Empirical scaling relations of GMC properties have given us important hints in this regard, but their interpretation can be affected by observational biases \citep{Ballesteros-Paredes&MacLow2002}. 

Three important scaling relations were first described by \citet{Larson1981}: first, a power-law relation between the velocity dispersion $\sigma$ (as measured from the linewidth) and the size $R$ of a CO-emitting region; second, an almost one-to-one relation between the observed masses and inferred virial masses of GMCs; and, third, a constant mass surface density of the analysed clouds. 

The validity of these scaling relations has been challenged both in the local environment and in extragalactic targets with different environmental conditions. Most observations do find a correlation between the size and the linewidth, but there seems to be no agreement for the power-law exponent and all observations find large uncertainties in the slope \citep[][]{Imara+2020, Duarte-Cabral+2020}. Moreover the third Larson relation is likely to be an artefact produced by the limited dynamical range of early observations of column density, and the sample of observed clouds being located in similar environmental conditions. Several studies have indeed confirmed that surface densities of GMCs can span over two orders of magnitude \citep{Heyer+2009, Hughes+2010, Hughes+2013, Leroy+2015, Duarte-Cabral+2020} though other surveys still find confirmation of the third Larson relation \citep{Lombardi&Alves&Lada2010}. 

If we assume that GMCs are virialised objects (i.e.\ that the second Larson relation holds), then the first Larson relation $\sigma \propto R^{1/2}$ naturally follows, which implies that the first two Larson relations are in fact not independent. However, if we acknowledge that the mass surface density varies among GMCs, then the constant of proportionality of the linewidth-size relation has a dependency on the surface density $\Sigma$. This correction to the first Larson law is summarised in the Heyer \citeyearpar{Heyer+2009} relation
\begin{equation} \sigma / R^{1/2} \propto \Sigma^{1/2}.\end{equation}
While the dependence on $\Sigma$ is generally acknowledged, GMCs seem to lie above the line predicted for clouds in self-gravitating virial equilibrium. Clouds in free-fall collapse naturally develop velocity dispersions that are close to, but slightly larger than, the viral equilibrium values \citep{Ballesteros-Paredes+2011}, and these velocities have the same functional dependence on $\Sigma$. Therefore, while clouds internal motions are normally assumed to oppose gravitational collapse, this interpretation is not unique, as inward collapse motions give a similar signature. The set of molecular clouds formed in the self-gravitating MHD simulations of \citet{Ibanez+2016}, for instance, are mainly gravitationally bound but still recover the observed velocity dispersion relations.

We also have to consider that GMCs are not isolated objects, and their environment could play an important role in confining the clouds. The tendency of observed clouds to be mainly gravitationally unbound when a virial analysis is performed could be explained by an external pressure confining force which, when considered, retrieves virial stability \citep{Field+2011, duarte-cabral&dobbs2017}. This would give the size-linewidth scaling relation an additional dependency on external pressure and prove the importance of galactic environment for GMC dynamics.

But again, it is not obvious that virialised structures should be expected from a turbulent gas flow. For instance, the energies of colliding streams that generate molecular structures \citep{Ballesteros-Paredes+1999} may be unrelated to the gravitational energy of the gas involved, since the former will be driven by external galactic phenomena. However, given the short dynamical times of the ISM these structures could virialise quickly. Moreover the energies within the turbulent cascade are not completely unrelated to the mass of GMCs but to some extent coupled through feedback processes.

Conditions leading to cloud formation could in general promote structures that are preferentially gravitationally unbound \citep{Dobbs+2011}. This could justify the low star formation efficiency of the dense ISM \citep{Kauffmann+2013} without having to invoke internal feedback processes to disperse the cloud \citep[e.g.][]{Federrath2015}.

The universality of GMC properties and their environmental dependence is crucial to resolve these controversies and to understand the dynamics regulating the molecular gas. Retrieving cloud statistics for extragalactic sources is technically challenging, but now achievable in the ALMA era. The first Larson relation in its original form is retrieved in NGC300 \citep{Faesi+2018}, but in the case of M51, no or only a weak size-linewidth correlation was found with large scatter \citep{Colombo+2014}. \citet{Hughes+2013} also find differences in cloud properties among M33, M51 and the LMC. These studies suggest that GMC properties are unlikely to be universal, and must hide a more complex dependence on other factors.  

In the case of the Milky Way we now have a set of excellent molecular gas tracer surveys (e.g.\ SEDIGISM, \citealt{Schuller+2017}; CHIMPS, \citealt{Rigby+2019}; COHRS, \citealt{Dempsey+2013}) but no agreement is reached for the first Larson and the Heyer relation. The trends are observed, but the scatter is large, and different surveys reach different conclusions for the exponents. Within the disc no strong variations of GMC properties in relation to the position in the disc are observed \citep[][]{Duarte-Cabral+2020}. 

In external galaxies, on the other hand, there can be significant differences depending on the positions of the clouds. \citet{Braine+2018} find a radial dependence of properties in M33, confirmed by dedicated simulations \citep{Dobbs+2019}, and link the variation to global galactic properties such as mass surface density rather than to local feedback processes. In contrast, differences found between arm and inter-arm clouds in M51 are often attributed to stellar feedback and the presence of galactic spirals \citep{Colombo+2014}. The centres of galaxies seem to be a particularly interesting location for cloud dynamics.

\citet{Sun+2020} find that in 70 nearby galaxies the GMCs in the central regions (and in particular in barred galaxies) have higher velocity dispersion. They also see a moderate difference in surface density, velocity dispersion, turbulent pressure and virial parameters between arm and inter-arm clouds, but the scatter is large. In our own Galaxy, clouds in the centre seem to exhibit larger line-widths suggesting larger turbulent driving which \citet{shetty+2012} suggest could be due to larger star formation, environmental densities and pressures with respect to the local ISM.

Interactions and galaxy mergers could also potentially affect the properties of the cold molecular gas. \citet{Pettitt+2018} find that clouds in a simulated interacting galaxy are generally more massive and have higher velocity dispersion than in an isolated one. As the tidal interaction induces the formation of spiral arms, the smaller clumps seen in the isolated disc cluster when they enter the arm to form larger mainly unbound clouds. 

One major barrier to disentangling the contrasting results of different GMC studies is the definition of a GMC itself. Thanks to the self-shielding property of H$_2$, the molecular phase of the ISM organises itself into structures with relatively sharp boundaries in terms of density, temperature and chemical state. This has led many authors to study the molecular ISM in terms of distinct clouds. Of course this is an oversimplification, and the real molecular phase exhibits a rich and complex morphology as the GMCs interact, merge, aggregate and dissipate.

Nevertheless, it remains useful to be able to partition the molecular phase into discrete structures. Many different segmentation schemes have been developed, each having their own strengths and weaknesses \citep[e.g.][]{ Stutzki&Guesten1990, Williams+1995, Rosolowsky&Leroy2006}. One must bear in mind that the definition of such structures cannot be unique and universal due to the complexity of the molecular ISM. Defining cloud boundaries can vary with the algorithmic approach used and can be artificial when trying to examine the global structure of the molecular phase. Scaling relations such as the first Larson relation in its original form are scale invariant and as such less sensitive to cloud definition. But properties like the cloud mass or virial parameter strongly depend on the cloud definition, which could produce artefacts and spurious results. Even with a consistent use of a specific cloud identification method, bias could arise in different ways, for example because of different environmental conditions. These could lead to crowding, which is a difficult problem for any segmentation method. Moreover resolution is key for these schemes and different beam sizes would lead to different structure identifications.

This problem is acknowledged in the literature, and we argue that despite the usefulness of defining discrete objects, these studies need to be augmented with a more general method, such as an analysis of the hierarchical structure of the ISM that describes molecular gas properties as a function of iso-(column)density levels \citep[as suggested for instance by][]{Hughes+2013}. This would not put particular emphasis on special scales and be less dependent on resolution and environmental conditions. Together with a cloud finding method this would give the most descriptive view of the molecular phase. 

In this work, we aim to improve our general comprehension of GMCs by performing simulations of the molecular gas in an interacting M51-like galaxy and studying cloud statistics. We identify and study individual clouds but we also use a dendrogram analysis \citep{Rosolowsky+2008, Goodman+2009,Colombo+2015} to identify structure at all levels. We focus in particular on the variation of cloud population properties in different environments and on the comparison of an interacting to an isolated galaxy. In Section~\ref{sec:methods} we summarize the galaxy models we use from \citet{Tress+2019} and describe the dendrogram analysis and how we identify clouds. In Section~\ref{sec:props} we present the derived properties of our identified structures. We further discuss their dependence on galactic location and environment in Section~\ref{sec:gal}, discuss the implications of missing physics in our simulations in Section~\ref{sec:missingPhysics} and we summarize and conclude in Section~\ref{sec:summ}.
   
\section{Methods}
\label{sec:methods}

\begin{figure*}
	\includegraphics[width=\textwidth]{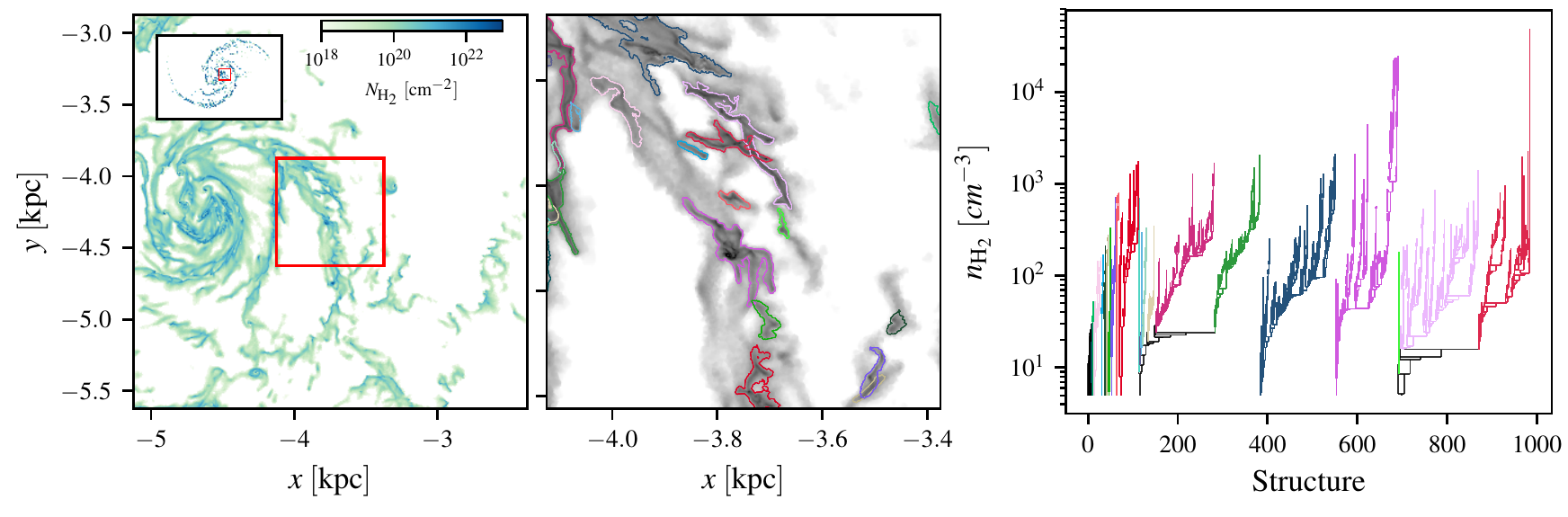}
    \caption{A region of the simulations in which we applied the cloud-finding method. In the left-hand panel we show the selected region ({\em red rectangle}) in the larger galactic context. The region of the left-hand panel is shown in reference to the whole galaxy in the small box in the top-left corner of the left panel. The dendrogram decomposition of this region is shown in the {\em right-hand panel}. {\sc scimes} then performs the segmentation and each different structure found is highlighted with a different colour. The location of these structures in the region studied is shown in the middle panel.}
    \label{fig:ScimesExample}
\end{figure*}

\subsection{Setup and simulations}
\label{subsec:setup}
An in-depth description of the setup and the simulation details can be found in \citet{Tress+2019}. Here we briefly summarise the most important features which are relevant for a clear and self-contained understanding of this manuscript.

The simulations were performed in order to study how the dense molecular phase of the ISM responds to galactic-scale events such as a galaxy interaction.
We took the M51 galaxy system as a template and our initial conditions were chosen so that at the end of the simulations we roughly reproduce the properties of this interacting galaxy. The model of the main galaxy comprises a dark matter halo, a stellar bulge and disc, and a gaseous disc. All these components and their mutual gravitational interactions are self-consistently evolved by the code throughout the simulation. The companion galaxy, on the other hand, is represented by a single massive collisionless particle.

We use the {\sc arepo} code \citep{Springel2010} to evolve the system in time, finding gravitational forces by solving the Poisson equation and, for the gas, solving the unmagnetised, hydrodynamic equations, including the energy equation.
We include the major physical ingredients thought responsible for shaping and controlling the life-cycle of GMCs. In particular, we include a non-equilibrium chemical network which is able to trace the hydrogen chemistry as well as a simple treatment for the formation and destruction of CO \citep{Glover+2012}. To do so we require information about the local non-ionizing UV interstellar radiation field that can photo-dissociate H$_2$ and CO. We assume a constant background radiation field and estimate the local shielding by computing for each cell the foreground column densities of the gas with the {\sc treecol} algorithm \citep{Clark+2012}. Radiative and chemical heating and cooling of the gas is followed as described in \citet{Clark+2019}.

Jeans unstable regions inside GMCs will gravitationally collapse leading to star formation. We abstract the late stages of collapse by employing accreting sink particles that are described in detail in \citet{Tress+2019}. Briefly, on each hydrodynamical timestep, we flag as candidates for sink particle formation all active cells\footnote{By default, {\sc arepo} uses a hierarchical time-stepping scheme and so only a subset of cells are updated on any given timestep. Cells that are updated on the current timestep are termed active cells.} that are above a pre-chosen density threshold, taken in these simulations to be $\rho_{\rm th} = 10^{-21}$~g~cm$^{-3}$. In order to actually form a sink, however, the candidate cells must pass a series of additional checks: they must be at a local minimum in the gravitational potential, the gas surrounding them must be gravitationally bound and converging, and they must not lie within the accretion radius of an existing sink particle. Cells that pass all of these checks are converted to collisionless sink particles with the same mass and momentum. Cells that have densities $\rho>\rho_{\rm th}$ but that lie within the accretion radius of an existing sink particle cannot form new sink particles. Instead, we check to see whether the gas in the cell is gravitationally bound to the sink. If it is, we remove enough mass from the cell to reduce its density to $\rho_{\rm th}$ and add this mass to the sink. Cells that lie within the accretion radius of multiple sinks give their gas to the sink to which they are most strongly bound. In the simulations analyzed in this paper, we adopt a sink accretion radius $r_{\rm acc} = 2.5$~pc and use the same value for the gravitational softening length of the sink. The gravitational softening length of the gas cells is adjusted adaptively as described in \citet{Springel2010} so that it always roughly matches the cell size. 

\begin{table}
	\centering
	\caption{Average and median properties of the cloud population of the interacting simulation at $t=217$~Myr.}
	\label{tab:cloudProperties_average}
	\resizebox{\columnwidth}{!}{\begin{tabular}{cccccc}
		\hline\hline
	    &	Mass (M$_\odot$) & Size (pc) & $\sigma_{\rm 1D}$ (km s$^{-1}$) & $\alpha_{\rm vir}$ & $j$ (km s$^{-1}$ pc) \\
		\hline
	    Average & $6.45 \times 10^4$  & $17.4$ & $5.09$ & $22.1$ & $1.13 \times 10^2$  \\
	    Median  & $1.99 \times 10^4$  & $16.0$ & $2.56$ & $6.46$ & $27.7$  \\
		\hline
	\end{tabular}}
\end{table}

We assume that $5$\% of the accreted gas is converted into stars. This mass is then used to populate a Kroupa initial mass function \citep{Kroupa2001} using the method described in \citet{Sormani+2017}. Based on the number of massive stars formed, we can attribute feedback coming from the sink particle, which represents a small young stellar cluster. We consider only SN feedback, neglecting ionization, stellar winds, or protostellar jets. At the end of the life-time of each massive star we create an SN event around the sink. This injects energy as well as returning the part of the sink mass that was not involved in the SF back into the ISM. 

In terms of resolution we set a base mass for the gas cells of $300$~M$_\odot$ but require that the Jeans length is always resolved by a minimum of four cells. This means that a cell will be refined if it has a mass greater than twice its base mass, or if the Jeans length is smaller than four times the cell diameter, whichever condition is more stringent. This grants us sub-parsec resolution inside of the GMCs and highest mass resolutions of about $10$~M$_\odot$ (see Figure 3 of \citealt{Tress+2019}).

The same physical setup, but with a sink particle formation density threshold of $\rho_{\rm th}=10^{-20}$~g~cm$^{-3}$ was successfully used in \citet{Tress+2020} and \citet{Sormani+2020} to study gas dynamics and star formation in the Central Molecular Zone of our Galaxy.

Along with the interacting M51-like galaxy, we also performed a simulation of the system in isolation. The same simulation was used to address the effect of the galactic encounter on the ISM properties by \citet{Tress+2019}. 

\subsection{Cloud identification}
\label{subsec:id}

\begin{figure*}
	\includegraphics[width=\textwidth]{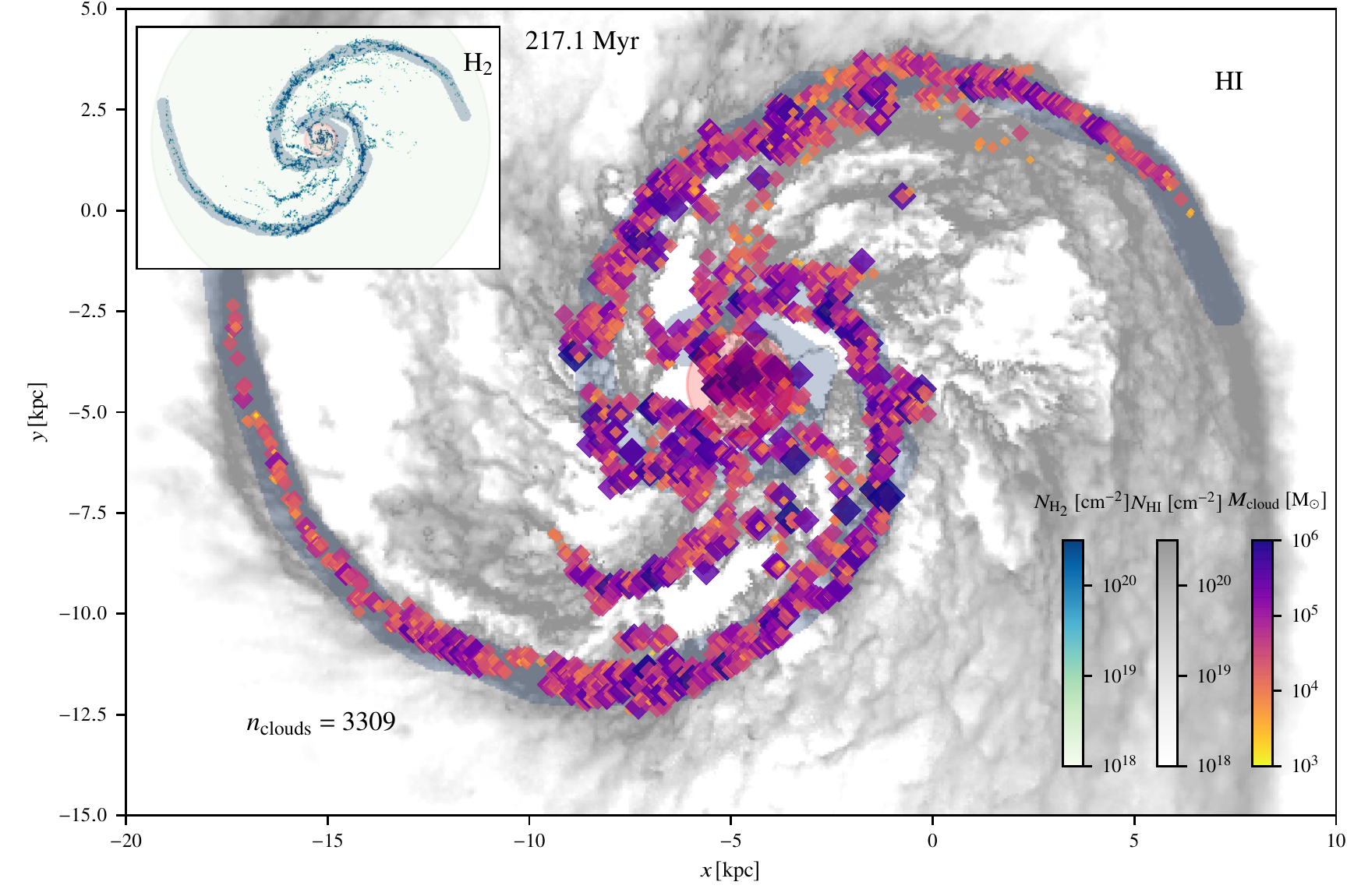}
        \caption{Positions of all the clouds found by the algorithm. The colour and the size of each marker is logarithmically related to the mass of the cloud. We also show the background HI column density ({\em gray-scale colour map}). On the {\em top left insert} we show the molecular hydrogen column density map. Clouds can be associated with spiral arms, ({\em dark blue shaded region}), with the inter-arm region ({\em unshaded region}) or with the nucleus ({\em red shaded region}).}
    \label{fig:CloudPositions}
\end{figure*}

\begin{figure}
	\includegraphics[width=\columnwidth]{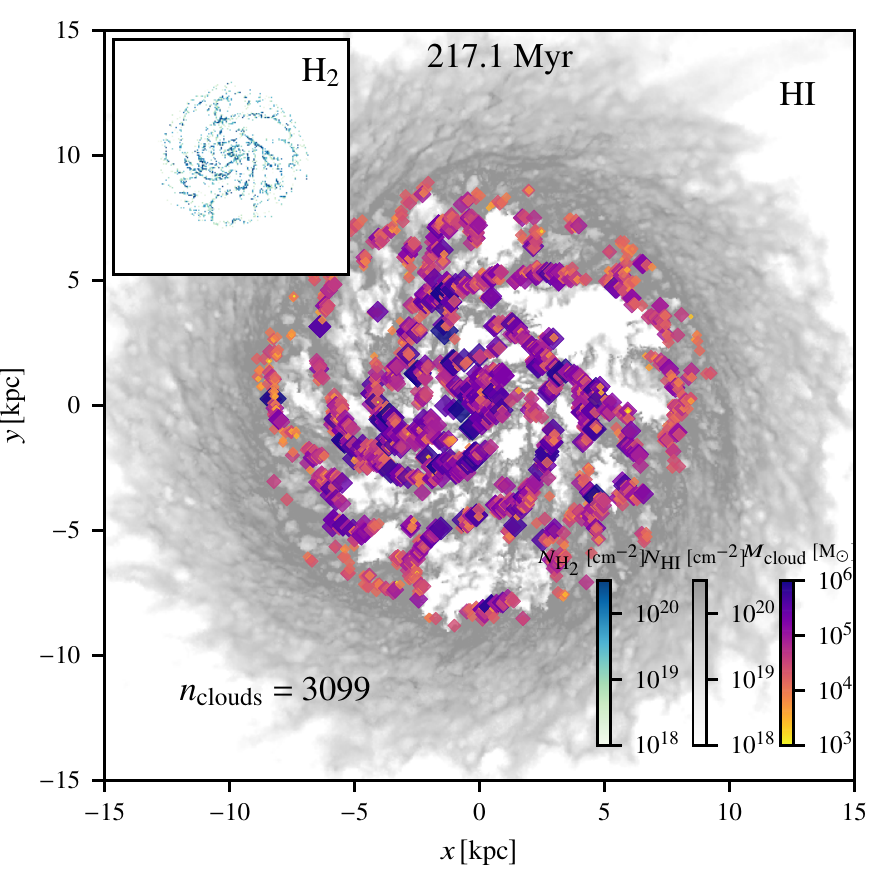}
        \caption{Same as Figure~\ref{fig:CloudPositions} but for the isolated galaxy.} 
    \label{fig:CloudPositions_iso}
\end{figure}

\begin{figure*}
    \begin{subfigure}{\textwidth}
    	\includegraphics{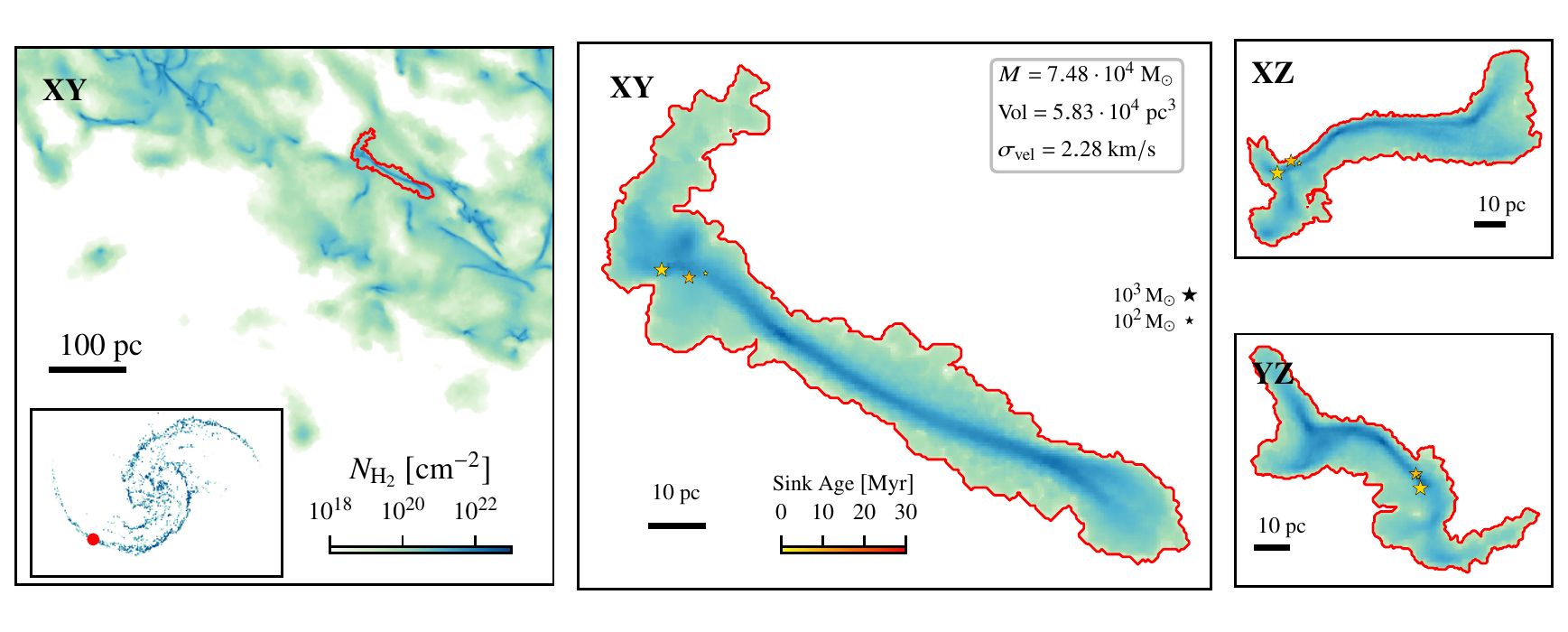}
    \end{subfigure}
    \begin{subfigure}{\textwidth}
    	\includegraphics{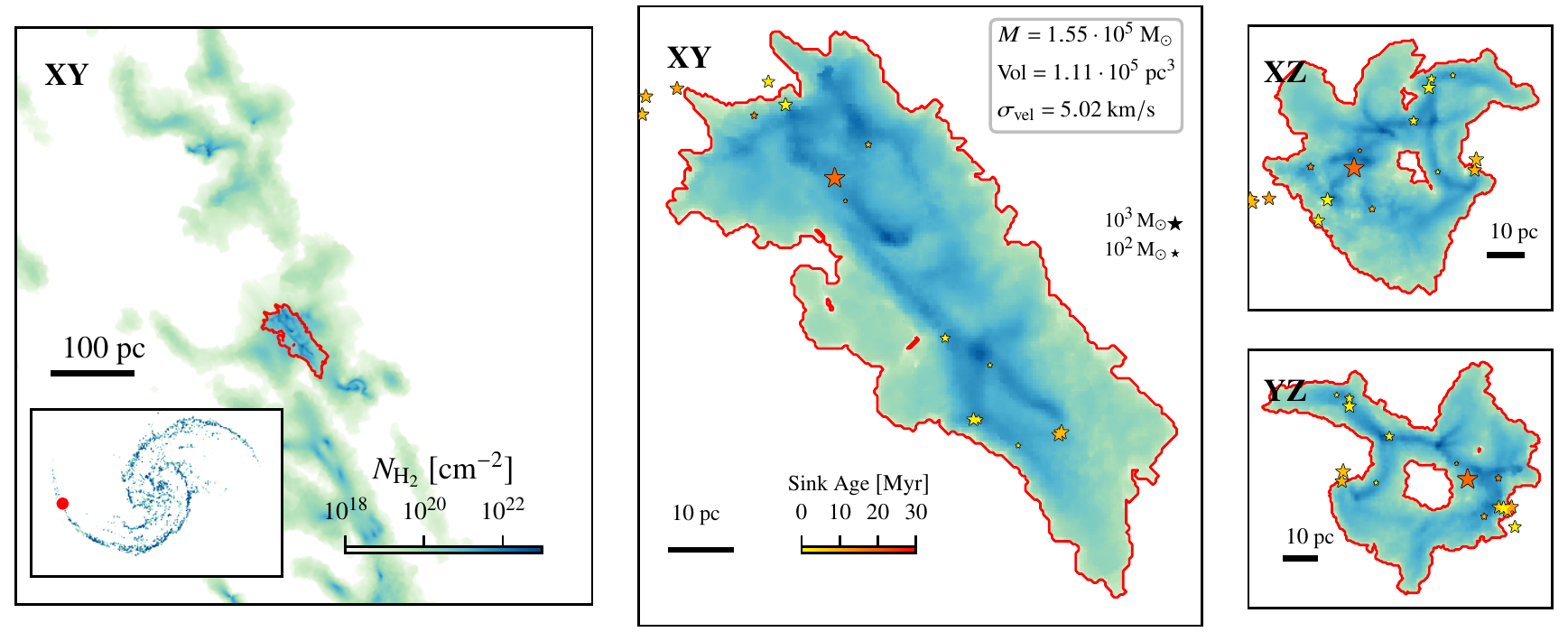}
    \end{subfigure}
    \begin{subfigure}{\textwidth}
    	\includegraphics{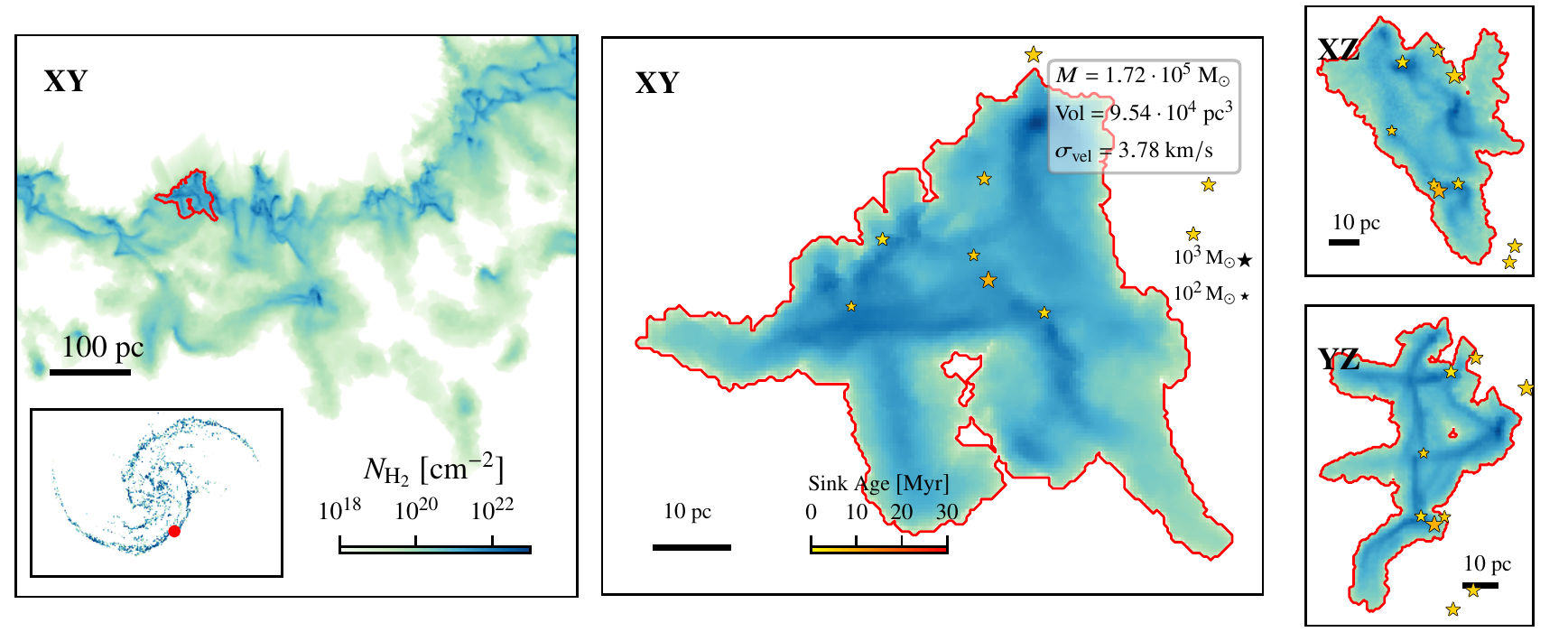}
    \end{subfigure}
    \caption{Examples of clouds identified by {\sc scimes}. On the left we show the H$_2$ column density map of the region from which the cloud was extracted. The inset shows the location of the region in the larger galactic context. The red isodensity contour identifies the cloud found by the algorithm. On the right we show the same cloud in the $XY$, $XZ$ and $YZ$ planes. We also indicate the locations of the sink particles using symbols colored by the sink age and with sizes related to the stellar mass.}
    \label{fig:CloudExample}
\end{figure*}
\begin{sidewaystable*}
  \resizebox{\textwidth}{!}{\begin{tabular}{cccccccccccc}\hline \hline
ID & $M_{\rm cloud}$ & Volume & $M_{\rm sink\, gas}$ & $\alpha_{\rm vir}$ & $(x, y, z)$ & $(v_x, v_y, v_z)$ & $(\sigma_x, \sigma_y, \sigma_z)$ & $(\sigma_{{\rm rot}, x}, \sigma_{{\rm rot}, y}, \sigma_{{\rm rot}, z})$ & $(j_x, j_y, j_z)$  & $\Omega$ & Location \\

& (M$_\odot$) & (pc$^3$) & (M$_\odot$) & & (kpc) & (km s$^{-1}$) & (km s$^{-1}$) & (km s$^{-1}$)  & (km s$^{-1}$ pc) & (Myr$^{-1}$) & \\\hline

$1320$ & $5.8 \cdot 10^3$ & $7.7 \cdot 10^3$ & -- & $50.4$ & $(-5.990, -8.738, 6.850)$ & $(196.1, -140.8, 73.3)$ & $(3.2, 4.3, 3.1)$ & $(0.8, 0.4, 1.8)$ & $(-23.4, 1.3, -8.8)$ & $2.9 \cdot 10^{-1}$ & I\\ 

$1755$ & $3.7 \cdot 10^4$ & $6.7 \cdot 10^4$ & $5.1 \cdot 10^3$ & $13.7$ & $(-4.861, -11.306, 6.851)$ & $(218.8, 9.4, 78.8)$ & $(3.8, 3.3, 3.3)$ & $(1.2, 1.3, 2.1)$ & $(-34.8, 23.6, 47.2)$ & $4.7 \cdot 10^{-1}$ & I\\ 

$1459$ & $2.4 \cdot 10^4$ & $2.7 \cdot 10^4$ & $4.7 \cdot 10^3$ & $9.5$ & $(-5.666, 1.382, 7.330)$ & $(-179.8, -128.6, 52.4)$ & $(1.9, 3.3, 3.0)$ & $(1.2, 0.1, 1.3)$ & $(10.9, 24.6, 11.9)$ & $5.0 \cdot 10^{-1}$ & A\\ 

$2152$ & $7.0 \cdot 10^4$ & $3.5 \cdot 10^4$ & -- & $45.3$ & $(-3.711, -9.993, 6.880)$ & $(207.1, 47.1, 100.7)$ & $(10.0, 10.5, 6.3)$ & $(5.8, 6.5, 2.0)$ & $(-101.5, 20.2, 193.7)$ & $1.7 \cdot 10^{-1}$ & A\\ 

$655$ & $5.2 \cdot 10^4$ & $3.3 \cdot 10^4$ & -- & $7.3$ & $(-8.100, -9.641, 6.803)$ & $(141.3, -197.2, 68.5)$ & $(2.7, 3.7, 3.1)$ & $(0.7, 0.9, 3.3)$ & $(115.9, -70.9, 6.2)$ & $2.2 \cdot 10^{-1}$ & I\\ 

$1226$ & $4.8 \cdot 10^4$ & $4.2 \cdot 10^4$ & $3.3 \cdot 10^4$ & $2.5$ & $(-6.331, 0.895, 7.386)$ & $(-172.6, -159.1, 52.5)$ & $(2.8, 2.2, 1.6)$ & $(1.6, 0.9, 1.0)$ & $(-11.6, 37.6, -49.2)$ & $7.8 \cdot 10^{-1}$ & A\\ 

$184$ & $9.7 \cdot 10^3$ & $9.9 \cdot 10^3$ & -- & $2.1$ & $(-12.684, -10.532, 6.033)$ & $(126.5, -215.3, 99.2)$ & $(0.5, 0.8, 1.2)$ & $(0.1, 0.3, 0.3)$ & $(-2.2, -1.7, 2.0)$ & $8.7 \cdot 10^{-1}$ & A\\ 

$2504$ & $1.3 \cdot 10^4$ & $9.1 \cdot 10^3$ & -- & $38.4$ & $(-2.335, -2.124, 7.276)$ & $(-143.2, 181.3, 96.6)$ & $(3.9, 4.9, 4.7)$ & $(2.4, 2.6, 1.8)$ & $(-9.8, 15.0, 33.5)$ & $1.3 \cdot 10^{-1}$ & A\\ 

$1634$ & $2.5 \cdot 10^4$ & $2.3 \cdot 10^4$ & $2.3 \cdot 10^4$ & $6.5$ & $(-5.255, 1.280, 7.768)$ & $(-152.1, -90.9, 49.3)$ & $(2.9, 2.9, 3.3)$ & $(0.9, 0.7, 1.3)$ & $(-10.3, 14.1, 0.7)$ & $3.6 \cdot 10^{-1}$ & A\\ 

$1651$ & $2.7 \cdot 10^4$ & $2.3 \cdot 10^4$ & -- & $1.0$ & $(-5.185, 1.347, 7.729)$ & $(-156.6, -91.8, 43.6)$ & $(1.0, 0.7, 1.0)$ & $(0.4, 0.3, 0.1)$ & $(0.2, 2.3, -7.4)$ & $1.4 \cdot 10^{0}$ & A\\ 

$1887$ & $1.4 \cdot 10^4$ & $1.3 \cdot 10^4$ & $4.1 \cdot 10^3$ & $4.0$ & $(-4.382, -10.684, 6.766)$ & $(212.0, 61.7, 96.0)$ & $(1.1, 0.9, 2.4)$ & $(0.3, 0.6, 1.3)$ & $(6.7, 14.4, -6.5)$ & $4.8 \cdot 10^{-1}$ & A\\ 

$524$ & $3.2 \cdot 10^5$ & $2.0 \cdot 10^5$ & $1.2 \cdot 10^5$ & $3.8$ & $(-8.517, -12.011, 6.462)$ & $(179.2, -98.0, 74.0)$ & $(6.2, 5.4, 2.4)$ & $(1.2, 3.2, 0.9)$ & $(-1.1, -7.7, 212.2)$ & $8.1 \cdot 10^{-1}$ & A\\ 

$1664$ & $2.9 \cdot 10^4$ & $2.1 \cdot 10^4$ & $3.7 \cdot 10^3$ & $0.8$ & $(-5.132, 1.379, 7.772)$ & $(-157.1, -87.4, 44.4)$ & $(0.7, 1.0, 1.1)$ & $(0.1, 0.3, 0.2)$ & $(-4.6, -1.2, 5.7)$ & $9.5 \cdot 10^{-1}$ & A\\ 

$2173$ & $7.8 \cdot 10^4$ & $6.5 \cdot 10^4$ & -- & $7.6$ & $(-3.661, -7.873, 7.036)$ & $(246.3, -19.9, 84.1)$ & $(3.3, 4.5, 2.5)$ & $(2.6, 1.5, 3.3)$ & $(-96.9, -92.1, 67.1)$ & $4.4 \cdot 10^{-1}$ & I\\ 

$826$ & $3.4 \cdot 10^3$ & $3.4 \cdot 10^3$ & -- & $19.0$ & $(-7.567, -11.220, 6.470)$ & $(219.4, -61.1, 117.8)$ & $(1.9, 2.6, 0.8)$ & $(0.4, 0.3, 0.2)$ & $(-1.0, -2.5, 1.5)$ & $6.6 \cdot 10^{-1}$ & I\\ 

$1378$ & $1.3 \cdot 10^4$ & $1.3 \cdot 10^4$ & -- & $7.5$ & $(-5.857, -8.609, 6.883)$ & $(204.8, -135.4, 77.8)$ & $(1.6, 2.6, 1.0)$ & $(0.1, 0.1, 0.1)$ & $(0.8, 0.4, 1.8)$ & $1.6 \cdot 10^{0}$ & I\\ 

$464$ & $1.6 \cdot 10^4$ & $1.3 \cdot 10^4$ & -- & $24.1$ & $(-8.854, -2.928, 6.884)$ & $(-83.3, -274.9, 72.2)$ & $(4.5, 2.6, 3.9)$ & $(0.7, 0.2, 2.3)$ & $(-7.2, 34.6, -3.4)$ & $3.6 \cdot 10^{-1}$ & I\\ 

$243$ & $2.5 \cdot 10^4$ & $2.3 \cdot 10^4$ & -- & $12.5$ & $(-12.152, -10.708, 5.971)$ & $(118.4, -182.5, 76.3)$ & $(2.6, 3.4, 3.2)$ & $(1.7, 0.8, 0.5)$ & $(-20.9, 35.2, 19.3)$ & $6.6 \cdot 10^{-1}$ & A\\ 

$3121$ & $3.7 \cdot 10^4$ & $3.6 \cdot 10^4$ & $1.1 \cdot 10^4$ & $3.3$ & $(1.228, 3.166, 8.774)$ & $(-105.1, 109.8, 73.7)$ & $(1.4, 1.6, 2.7)$ & $(1.4, 1.5, 1.5)$ & $(30.7, -16.0, -17.8)$ & $2.3 \cdot 10^{-1}$ & A\\ 

$2857$ & $2.5 \cdot 10^5$ & $1.8 \cdot 10^5$ & $7.7 \cdot 10^4$ & $3.6$ & $(-1.200, 3.117, 8.541)$ & $(-152.3, 35.9, 72.4)$ & $(3.1, 5.2, 4.2)$ & $(0.3, 3.6, 1.3)$ & $(-230.0, -61.5, 153.2)$ & $7.5 \cdot 10^{-1}$ & A\\ 

$3228$ & $1.1 \cdot 10^4$ & $1.1 \cdot 10^4$ & -- & $16.7$ & $(3.253, 2.583, 9.165)$ & $(-72.4, 163.3, 89.3)$ & $(1.6, 3.0, 3.2)$ & $(1.2, 0.7, 0.5)$ & $(16.1, 21.1, 13.2)$ & $8.0 \cdot 10^{-1}$ & A\\ 

$360$ & $3.6 \cdot 10^3$ & $2.0 \cdot 10^3$ & -- & $21.3$ & $(-10.205, -11.180, 6.065)$ & $(191.6, -136.6, 99.3)$ & $(2.1, 1.7, 2.9)$ & $(0.3, 1.9, 1.1)$ & $(7.8, 1.6, -7.6)$ & $1.3 \cdot 10^{-1}$ & A\\ 

$307$ & $7.9 \cdot 10^3$ & $1.4 \cdot 10^4$ & -- & $3.6$ & $(-11.512, -11.342, 6.458)$ & $(152.2, -196.5, 89.3)$ & $(1.1, 1.0, 0.8)$ & $(0.1, 0.4, 0.4)$ & $(5.0, 0.6, -0.1)$ & $7.8 \cdot 10^{-1}$ & A\\ 

$2008$ & $1.5 \cdot 10^4$ & $1.7 \cdot 10^4$ & -- & $36.7$ & $(-3.997, 2.281, 7.716)$ & $(-193.2, -46.5, 63.6)$ & $(1.4, 6.7, 2.5)$ & $(1.1, 1.2, 0.9)$ & $(8.0, -1.8, 17.9)$ & $3.2 \cdot 10^{-1}$ & A\\ 

$464$ & $1.6 \cdot 10^4$ & $1.3 \cdot 10^4$ & -- & $24.1$ & $(-8.854, -2.928, 6.884)$ & $(-83.3, -274.9, 72.2)$ & $(4.5, 2.6, 3.9)$ & $(0.7, 0.2, 2.3)$ & $(-7.2, 34.6, -3.4)$ & $3.6 \cdot 10^{-1}$ & I\\ 

$1174$ & $3.8 \cdot 10^4$ & $2.2 \cdot 10^4$ & -- & $1.8$ & $(-6.483, 0.556, 7.342)$ & $(-145.3, -180.3, 46.5)$ & $(1.3, 1.6, 1.5)$ & $(1.1, 0.3, 0.6)$ & $(11.2, -13.5, 30.3)$ & $5.7 \cdot 10^{-1}$ & A\\ 

$2636$ & $1.0 \cdot 10^4$ & $1.2 \cdot 10^4$ & -- & $14.4$ & $(-1.836, -3.859, 7.087)$ & $(-22.2, 153.3, 82.4)$ & $(2.0, 3.2, 1.3)$ & $(0.3, 0.2, 0.5)$ & $(5.6, 2.4, -2.1)$ & $8.7 \cdot 10^{-1}$ & A\\ 

$357$ & $2.7 \cdot 10^4$ & $2.5 \cdot 10^4$ & -- & $7.0$ & $(-10.320, -11.178, 5.924)$ & $(184.9, -129.5, 102.2)$ & $(0.9, 3.2, 2.4)$ & $(0.8, 2.2, 1.6)$ & $(21.5, 21.8, 14.0)$ & $2.2 \cdot 10^{-1}$ & A\\ 

$544$ & $3.2 \cdot 10^4$ & $2.4 \cdot 10^4$ & $1.6 \cdot 10^4$ & $2.0$ & $(-8.447, -9.545, 6.876)$ & $(130.8, -228.7, 76.5)$ & $(1.9, 1.6, 1.7)$ & $(0.9, 0.7, 1.0)$ & $(-12.7, -13.3, 23.3)$ & $6.0 \cdot 10^{-1}$ & I\\ 

$1783$ & $1.4 \cdot 10^4$ & $2.3 \cdot 10^4$ & -- & $23.0$ & $(-4.724, -11.095, 6.835)$ & $(193.0, 22.8, 52.4)$ & $(3.5, 1.3, 4.0)$ & $(0.4, 0.3, 0.2)$ & $(-2.1, -8.9, 2.8)$ & $1.1 \cdot 10^{0}$ & A\\ 

$1896$ & $1.4 \cdot 10^3$ & $1.1 \cdot 10^2$ & -- & $93.6$ & $(-4.326, -7.866, 6.956)$ & $(238.8, -38.9, 62.5)$ & $(4.8, 4.8, 4.8)$ & $(3.6, 1.0, 3.0)$ & $(0.9, -7.0, -0.1)$ & $1.8 \cdot 10^{-2}$ & I\\ 

$1083$ & $5.3 \cdot 10^5$ & $6.6 \cdot 10^5$ & $2.3 \cdot 10^4$ & $39.9$ & $(-6.701, -6.396, 6.848)$ & $(189.5, -177.3, 98.3)$ & $(11.8, 22.2, 4.5)$ & $(5.6, 12.7, 2.2)$ & $(151.9, -269.1, 1815.6)$ & $5.9 \cdot 10^{-1}$ & A\\ 

$1495$ & $3.0 \cdot 10^4$ & $5.6 \cdot 10^4$ & -- & $20.8$ & $(-5.602, -3.857, 7.127)$ & $(-46.7, -207.7, 59.6)$ & $(4.6, 4.0, 2.3)$ & $(2.3, 1.1, 0.6)$ & $(15.3, -7.1, 85.2)$ & $5.3 \cdot 10^{-1}$ & C\\ 

$2332$ & $1.1 \cdot 10^4$ & $1.2 \cdot 10^4$ & -- & $20.6$ & $(-3.099, -5.857, 7.140)$ & $(52.8, 93.2, 83.5)$ & $(4.8, 1.5, 0.8)$ & $(0.4, 0.7, 0.3)$ & $(0.4, -2.9, 10.5)$ & $6.7 \cdot 10^{-1}$ & A\\ 

$1650$ & $1.9 \cdot 10^4$ & $1.2 \cdot 10^4$ & -- & $21.2$ & $(-5.188, -10.636, 6.648)$ & $(210.2, 55.8, 126.1)$ & $(2.3, 6.2, 1.7)$ & $(2.2, 2.5, 0.4)$ & $(5.8, 4.2, -42.4)$ & $2.3 \cdot 10^{-1}$ & A\\ 

$1092$ & $5.2 \cdot 10^4$ & $2.8 \cdot 10^4$ & $3.3 \cdot 10^4$ & $10.0$ & $(-6.681, -3.248, 7.046)$ & $(-54.0, -212.0, 51.7)$ & $(4.4, 5.4, 4.9)$ & $(1.2, 1.7, 1.3)$ & $(12.0, -28.9, -51.0)$ & $5.7 \cdot 10^{-1}$ & A\\ 

$146$ & $1.0 \cdot 10^4$ & $1.4 \cdot 10^4$ & -- & $37.8$ & $(-13.100, -10.330, 6.136)$ & $(113.5, -234.8, 96.2)$ & $(4.9, 1.4, 3.9)$ & $(2.9, 1.4, 2.4)$ & $(-8.5, -33.9, -5.8)$ & $1.3 \cdot 10^{-1}$ & A\\ 

$7$ & $8.7 \cdot 10^3$ & $8.3 \cdot 10^3$ & -- & $13.8$ & $(-17.271, -2.892, 4.895)$ & $(-49.6, -232.1, 80.4)$ & $(2.8, 2.4, 1.4)$ & $(0.9, 1.0, 0.4)$ & $(-4.9, 0.3, -10.3)$ & $3.5 \cdot 10^{-1}$ & A\\ 

$2668$ & $3.7 \cdot 10^4$ & $2.1 \cdot 10^4$ & -- & $5.1$ & $(-1.714, -8.157, 7.040)$ & $(165.5, 183.2, 87.0)$ & $(3.0, 1.7, 2.4)$ & $(1.4, 0.7, 1.3)$ & $(24.7, 19.4, 9.1)$ & $4.5 \cdot 10^{-1}$ & A\\ 

$1168$ & $2.9 \cdot 10^4$ & $1.3 \cdot 10^4$ & -- & $8.8$ & $(-6.491, -12.081, 6.616)$ & $(196.4, -60.3, 74.6)$ & $(1.1, 5.1, 1.0)$ & $(1.8, 3.6, 1.0)$ & $(11.6, 4.9, 36.9)$ & $1.4 \cdot 10^{-1}$ & A\\

\hline
\end{tabular}}
\caption{Properties of a sub-sample of the molecular clouds found in the interacting simulated galaxy. In order from left to right we show the clouds identification number, mass, volume, mass of gas which is trapped in sink particles contained within the cloud if present, virial parameter, centre of mass position, velocity of the centre of mass, velocity dispersion, velocity dispersion coming from rotation, specific angular momentum, angular velocity and the type of cloud where A, I, C indicate clouds associated with the arms, inter-arms and centre respectively. The complete catalogue is provided in machine readable form as supplemental material.}
\label{tab:cloudProperties_clouds}
\end{sidewaystable*}

To identify clouds in our simulations, we make use of the dendrogram-based scheme {\sc scimes} \citep{Colombo+2015}. In its most general form, a dendrogram is a tree diagram indicating the hierarchical relationship between objects. Dendrograms are used in many different fields of science, with their use in astronomy being popularised by \citet{Rosolowsky+2008} and \citet{Goodman+2009}. Here, we use them to represent the relationship between different isodensity contours in the molecular ISM. Local density maxima are identified as the leaves of the dendrogram (i.e.\ structures at the top of the tree that enclose no further substructures). Isodensity contours corresponding to lower density values enclose multiple leaves and are called branches of the dendrogram. To construct our dendrogram, we use a set of H$_2$ isodensity contours that starts at a minimum density of $n_{\rm H_2, min} = 1$~cm$^{-3}$ and that has a spacing of $\Delta n_{\rm H_2} = 5$~cm$^{-3}$ between contours. An example of the resulting dendrogram decomposition of a small sub-region of the simulation is shown in the right-hand panel of Figure~\ref{fig:ScimesExample}.

Given the dendrogram decomposition of the H$_{2}$ density field, we then use the {\sc scimes} algorithm to segment it into a set of discrete structures. {\sc scimes} is a spectral clustering technique that groups the leaves of the dendrogram into clusters according to their similarity, as assessed by a set of user-supplied similarity criteria. In the present case, we use the volume and the mass of structures in the dendrogram as input properties for {\sc scimes}. This means that in general if there is an abrupt change in mass and volume while walking the dendrogram, the code will identify this location in the graph as the point at which to perform the segmentation.

In the dendrogram shown in Figure~\ref{fig:ScimesExample}, the different clusters identified by {\sc scimes} are highlighted with different colours. We identify each of these clusters as a distinct molecular cloud. The locations of these clouds in the $x$-$y$ plane are shown in the middle panel of Figure~\ref{fig:ScimesExample}, overlaid on a projection of the H$_{2}$ column density in this sub-region of the galaxy. We see that the clouds identified by {\sc scimes} correspond to regions with high H$_{2}$ column densities, as one would expect. In addition, we also see that there is a spatially extended distribution of low column density H$_{2}$ surrounding many of the clouds that is not associated by {\sc scimes} with a particular molecular cloud. This mostly corresponds to cold, neutral atomic gas with a low but non-zero H$_{2}$ fraction. These envelopes surrounding GMCs were identified as being largely CO dark in analogous high resolution simulations \citep{Smith+2014}.

We tried to vary the parameters for the dendrogram construction and for {\sc scimes} in a given region of the simulated galaxy to test the sensitivity of the cloud properties against the choice of those parameters. We found that the structures identified and their mass distribution is relatively insensitive to small variations of the density spacing between contours for the dendrogram construction around our chosen value of $\Delta n_{\rm H_2} = 5$~cm$^{-3}$. If instead the spacing increases too much, the algorithm has a tendency to merge structures which were identified as separate with a finer spacing. Small variations in the minimum density generally have a small impact on the mass of some identified clouds as the algorithm is prone to include (or exclude) lower density contours. By adopting a too small minimum density, however, the algorithm will sometimes fail to properly segment clouds in the spiral arm and identify the entire arm as a single cloud. We found that a change in the properties used by {\sc scimes} to assess similarity of dendrogram branches (using only volume, or only mass) can result in a great difference in structures identified, but a visual inspection of the clouds found suggests that a combination of mass and volume comes closest to how a user would proceed with the segmentation by hand. We notice that the definition of clouds is used in this work only in relative terms, and as long as the definition is consistently used throughout the domain, a comparison of cloud properties between regions is justified. Caution is instead advised when comparing the results from studies that used different criteria (parameters) in their cloud identification. 

One complication in our cloud identification method is that both {\sc scimes} and the software used to construct the dendrogram ({\sc astrodendro}) are only able to operate using isodensity contours defined on a regular grid. We therefore have to regrid the {\sc arepo} output, which is defined on an unstructured Voronoi mesh, onto a regular 3D Cartesian mesh. In order to retain all of the details of the simulation in the high density gas, the grid size needs to be smaller than the smallest native resolution. Since the smallest cells have sizes below $0.1$~pc (see \citealt{Tress+2019}, Figure 3), this proved to be computationally impractical owing to the extremely large size of the resulting grid. We therefore compromised by using a grid cell size of $0.5$~pc, which is small enough to capture the structure of the molecular clouds, while requiring more than an order of magnitude less memory than a $0.1$~pc grid. In addition, rather than representing the entire galaxy using a single grid, we instead sub-divided it into a series of $(500 \; \rm{pc})^3$ regions which could then be processed serially using {\sc scimes}. In order to avoid missing clouds located close to the boundary of a region, we overlapped each region by 125~pc with each of its neighbouring regions. For each region we only retain clouds with centre of mass within the original $(500 \; \rm{pc})^3$ box. In this way we avoid double-counting. 

We store the total density, H$_2$ density, and the velocity of the grid points associated with each cloud found with this method. We also determine and save the following properties: cloud ID, total gas mass, volume, mass of gas in any sink particles contained within the cloud, virial parameter, position and velocity of the centre of mass, velocity dispersion, velocity dispersion arising from rotation, specific angular momentum, and angular velocity. For a definition of these quantities see the relevant sections.

Molecular hydrogen is typically undetectable in real galaxies, with CO emission being the most widely used observational proxy for it. However, we do not use CO to identify our clouds and instead rely on the actual H$_2$ densities. There are two main reasons for this choice. First, although our resolution is extremely high by the standards of galactic-scale simulations, it is still not high enough to yield numerically converged values for the CO distribution. At typical GMC densities, most of the cells in the simulation have sizes of $\sim 0.5$~pc or larger, roughly an order of magnitude larger than the value of $\sim 0.05$~pc that \citet{Joshi+2019} find is necessary to obtain fully converged values for the CO distribution in simulations of turbulent molecular clouds. Second, our primary interest in this study is the morphology and dynamics of the entire molecular phase, i.e.\ all of the gas located in H$_{2}$ dominated regions, rather than just the subset of it which is rich in CO. In future work, we intend to compare the properties of clouds identified using CO emission with the properties of clouds identified using H$_{2}$ densities. However, this lies outside of the scope of our current study.

We also do not convolve the cubes with a Gaussian beam but instead use the native resolution to find structures. This limits any direct comparison to observations but gives us insight into the actual properties of the molecular gas.

\subsection{Cloud catalogue}
\label{sec:cloudCatalogue}
We compiled a cloud catalogue for the interacting simulation at $t=217$~Myr. The choice of this snapshot was rather arbitrary, but we also took samples of clouds at different times in the simulation and we do not find substantial difference in our results (see Appendix \ref{sec:differentTimes}). To see what the role of the galaxy interaction is in determining cloud properties, we also performed the cloud search algorithm on the same galaxy in isolation at the same simulation time.

\begin{figure*}
    \begin{subfigure}{\columnwidth}
    	\includegraphics[width=\columnwidth]{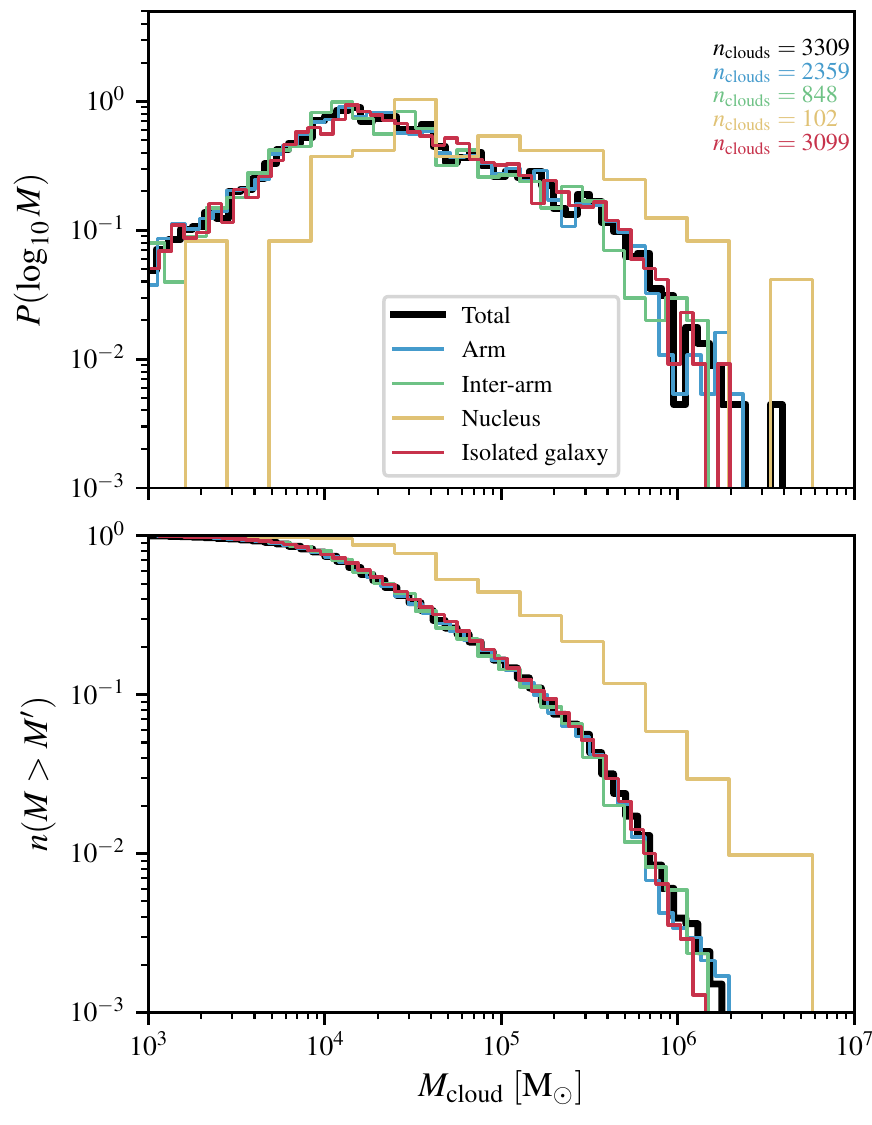}
    \end{subfigure}
    \begin{subfigure}{\columnwidth}
    	\includegraphics[width=\columnwidth]{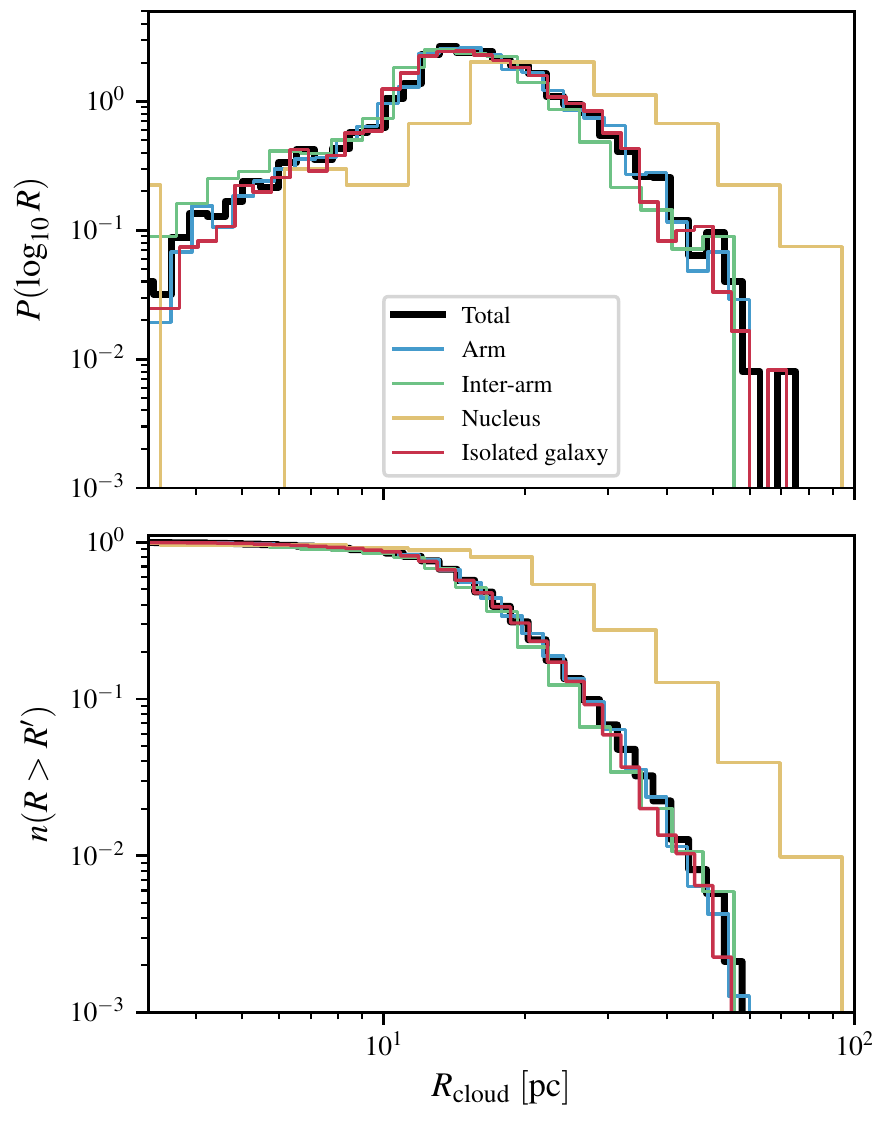}
    \end{subfigure}
    \caption{In the top panels we show the probability distribution function of the mass ({\em left}) and size ({\em right}) of the GMC population. The solid black line refers to the complete set of clouds, while the cloud population associated with spiral arms, inter-arm region, nucleus and the isolated galaxy are highlighted in different colours (see legend in the top panels). The size distribution of the simulated GMC catalogue $R_{\rm cloud}$ is computed from the clouds' volumes by assuming spherical shapes. The complementary cumulative distribution is shown in the bottom panels, where $n(M>M^{'})$ ($n(R>R^{'})$) denotes the fraction of clouds with mass (size) greater than a given value.}
    \label{fig:MassDistribution}
\end{figure*}

Figure~\ref{fig:CloudPositions} shows the positions of all clouds identified by our algorithm in the interacting galaxy while Figure~\ref{fig:CloudPositions_iso} shows the isolated case. A total of $3309$ and $3099$ clouds were identified in the interacting and isolated galaxy respectively. We summarise the global properties of the cloud population for the interacting simulation in Table \ref{tab:cloudProperties_average}.

Based on their positions, the clouds in the interacting simulation were assigned either to the spiral arms, to the inter-arm regions, or to the centre. The centre is defined as the area with $R<1.3$~kpc. To define the spiral arm region we perform a Fourier transformation on the H$_2$ column density in different radial bins. A similar method was used by \citet{Pettitt+2020} where only the $m=2$ mode was retained to determine the position of the arms. Due to the more complex structure of the spirals here, we retain also higher harmonics in order to be able to better trace the density peaks. The $m=2$ mode is dominant in the outer parts of the disc but some radial bins exhibit a multi-arm structure. For these bins we then only consider the continuation of the two-armed spiral pattern from the outer disc. Clouds are then assigned to the spiral arms if they lie within $500$~pc of this spiral arm spine (blue shaded region of Figure~\ref{fig:CloudPositions}). The largest agglomeration of molecular gas in the inter-arm region (around $(-6,-8)$~kpc in Figure~\ref{fig:CloudPositions}) could be interpreted as an additional arm, but has no counterpart in the stellar component. It is rather an over-density detaching from the spiral arm where the gas accumulated a few tens of megayears earlier, so we attribute it to the inter-arm region. 

The total gas mass in molecular clouds in the interacting galaxy is $M_{\rm tot} = 2.1 \times 10^8$~M$_\odot$ and the GMCs contain a total of $M_{\rm H_2, tot} = 9.9 \times 10^7$~M$_\odot$ of molecular hydrogen, not counting the mass trapped in sink particles. A total H$_2$ gas mass of $4.3 \times 10^7$~M$_\odot$ was not attributed to any GMC by the algorithm. 

In Table~\ref{tab:cloudProperties_clouds} we show the properties of a sub-sample of the clouds identified and in Figure~\ref{fig:CloudExample} some examples of the clouds at different locations in the interacting galaxy, as well as the position of the sink particles associated with the clouds. Even though we are far from resolving GMCs down to core scales, our resolution is high enough to show the complex and filamentary substructure of the clouds. A typical cloud of mass $2\times10^4$~M$_\odot$ is resolved by $\gtrsim 10^3$ {\sc arepo} cells, we therefore believe that a detailed study of their properties is appropriate and gives us important insight to their dynamics. 

\section{Cloud properties}
\label{sec:props}
\begin{figure*}
	\includegraphics[width=\textwidth]{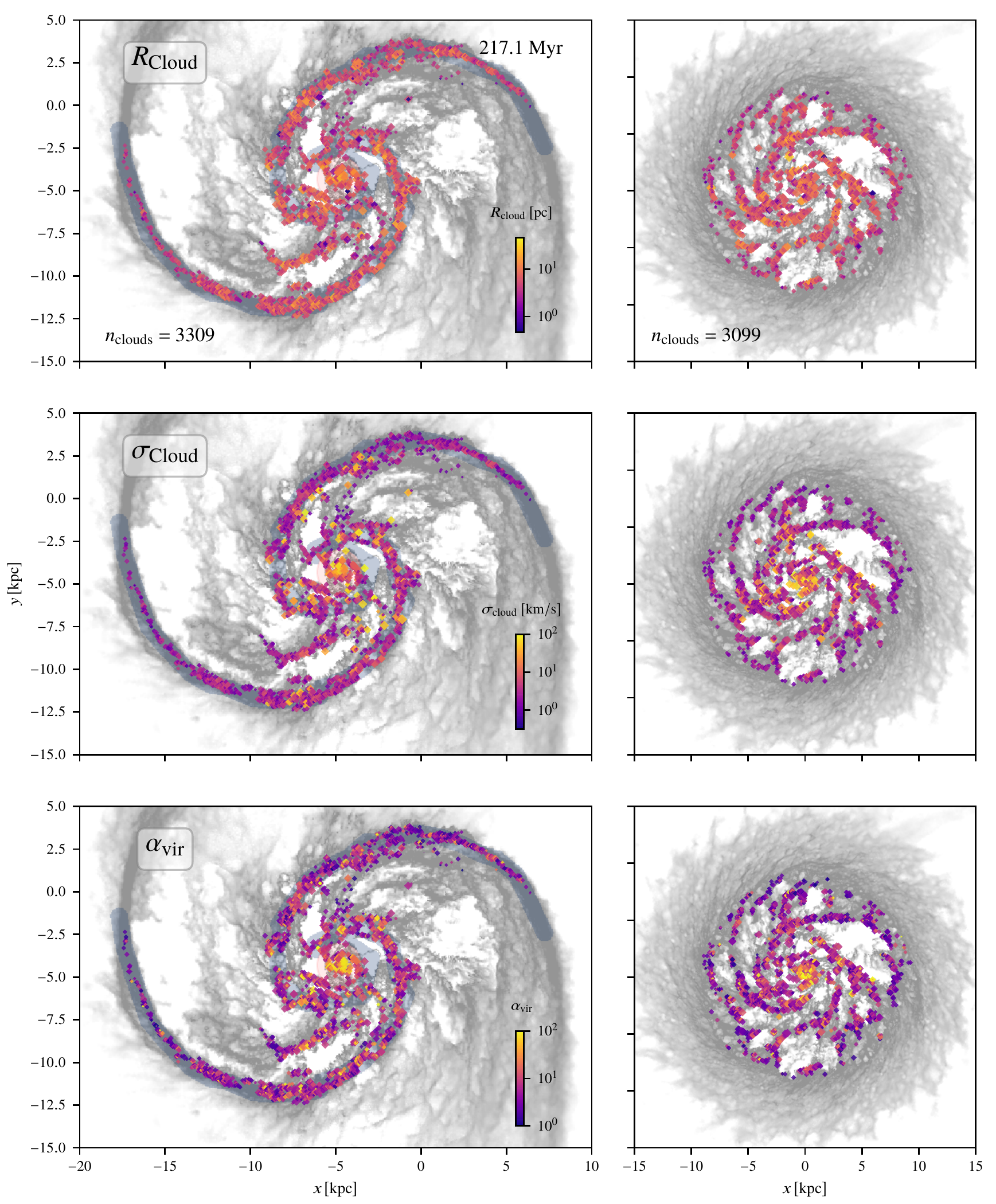}
    \caption{Identified clouds coloured based on their effective radius ({\em top panel}), their velocity dispersion ({\em middle panel}) and virial parameter ({\em bottom panel}). The marker size is related to the cloud mass as in Figure~\ref{fig:CloudPositions}. We show the interacting galaxy on the left and the isolated case at the same simulation time on the right. }
    \label{fig:cloudPropertiesMap}
\end{figure*}

\subsection{Masses}
\label{sec:Masses}

In Figure~\ref{fig:MassDistribution} we show the mass distribution of the clouds. To ensure that our analysis only considers well-resolved clouds, we exclude clouds found by {\sc scimes} with masses lower than $10^{3} \: {\rm M_{\odot}}$. The native mass resolution of the simulation at GMC densities is around 10~M$_{\odot}$, so all of the clouds identified that we keep for our analysis are resolved with around 100 or more {\sc arepo} resolution elements and the bulk of GMCs with more than $10^3$.

The most massive clouds found at this resolution have a mass of $\simeq 10^{6.5}$~M$_\odot$. This is considerably lower than the most massive structures identified by \citet{Colombo+2014} from CO observations of the M51 system whose clouds reach masses up to $10^{7.5}$~M$_\odot$. Our resolution is relatively high compared to their beam-size and the cloud finding algorithm is therefore able to pick out and segment smaller structures. 

The mass distribution in Figure~\ref{fig:MassDistribution} peaks at around $10^4$~M$_\odot$. This is a regime where the GMCs in our simulation are reasonably well resolved, so this peak is of considerable interest and may be an emerging property for simulations of clouds given the physics included in these models. However, the cloud-finding algorithm could potentially introduce bias here.

In Figure~\ref{fig:CloudPositions} we show the positions of the clouds coloured by their mass. Figures~\ref{fig:CloudPositions} and~\ref{fig:MassDistribution} show no evident difference between the cloud distribution of the arm and inter-arm regions. Clouds in the nucleus, however, are generally more massive and have a shallower mass distribution than the rest of the galaxy. Even the isolated galaxy exhibits a cloud population that appears indistinguishable from the interacting one.

\begin{figure*}
    \begin{subfigure}{\columnwidth}
    	\includegraphics[width=\columnwidth]{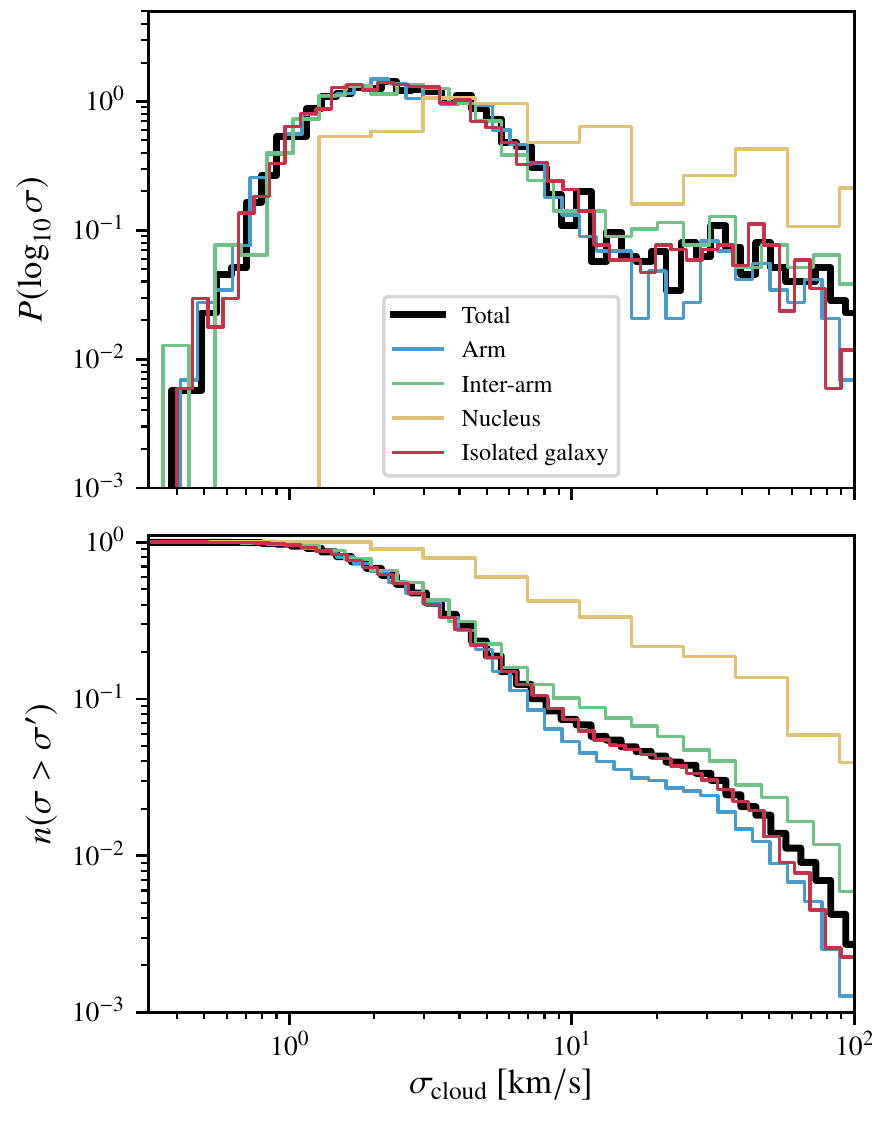}
    \end{subfigure}
    \begin{subfigure}{\columnwidth}
    	\includegraphics[width=\columnwidth]{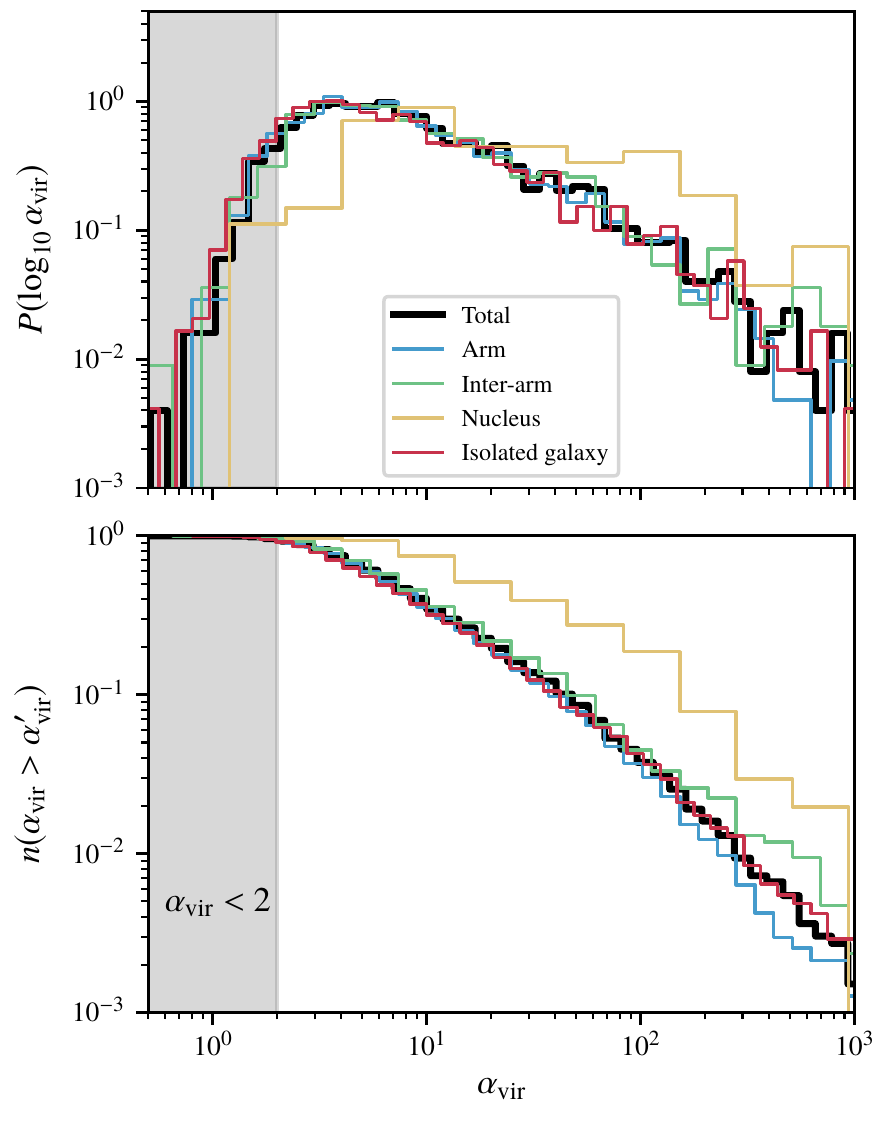}
    \end{subfigure}
    \caption{Velocity dispersion and virial parameter distributions of the simulated GMC catalog; $\sigma_{\rm cloud}$ is the 1D velocity dispersion of the cloud by assuming isotropic motions. The different cloud populations are depicted in different colours, consistent with our other figures. The cumulative distribution is shown in the bottom panel. The different cloud populations are depicted in different colours, consistent with our other figures. The cumulative distribution is shown in the bottom panel. The {\em gray band} is the region where $\alpha_{\rm vir} < 2$, where the virial analysis suggests that the structures are collapsing. Note that the majority of the clouds are close to the critical value, and gravitationally unbound GMCs are clearly favoured.}
    \label{fig:AlphaVirDistribution}
\end{figure*}

The absolute values of these distribution functions have to be viewed with some caution, as a considerable amount of gas could be trapped in sink particles. This is not included when computing the mass of GMCs as this would imply strong sub-grid assumptions on the thermodynamic and chemical state of this gas. The mass distribution is therefore most useful in relative terms, for comparing clouds in different regions within this framework.

\subsection{Sizes}
\label{sec:Sizes}

The effective radius of a specific cloud is computed by assuming the cloud to be spherical: 
\begin{equation}
    R_{\rm cloud} = \left( \frac{3}{4 \pi} V_{\rm cloud} \right)^{1/3}\,,
    \label{eq:R}
\end{equation}
where $V_{\rm cloud}$ is the total volume of the cloud. This is of course an over-simplification and in many instances might not represent the actual extension of the cloud, as GMCs can be represented by extremely elongated filaments or might contain holes in their distribution. Nonetheless this simple definition is useful to detect correlations given a statistically large sample of clouds.

From Figure~\ref{fig:MassDistribution} we see that the effective radius of our cloud population peaks at $\sim 20$~pc. There is no evident difference in size of clouds in the arm and clouds of the inter-arm region while GMCs in the nucleus clearly seem to belong to a different population. Here the large shearing forces are able to considerably stretch the clouds, thus producing a population whose clouds are generally a bit bigger. This would produce an imprint on shapes of clouds and we would expect larger aspect ratios of nuclear GMCs. On the other hand shear would be expected to efficiently disrupt large clouds, and we plan therefore to revisit the geometry of structures in an upcoming work to then tie it to the local shear and test these possibilities. We will see in the following sections that nuclear clouds stand out in velocity dispersion and virial parameter as well. This then fits in this picture as the high shear promotes higher velocity dispersion and could support larger clouds against collapse. 

In Figure~\ref{fig:cloudPropertiesMap} we show the spatial distribution of clouds in the galaxy coloured respectively by their size, velocity dispersion and virial parameter. The difference of central clouds compared to disc clouds is distinguishable here as well. 

\subsection{Velocity dispersion}
\label{sec:sigma}

\begin{figure}
	\includegraphics[width=0.94\columnwidth]{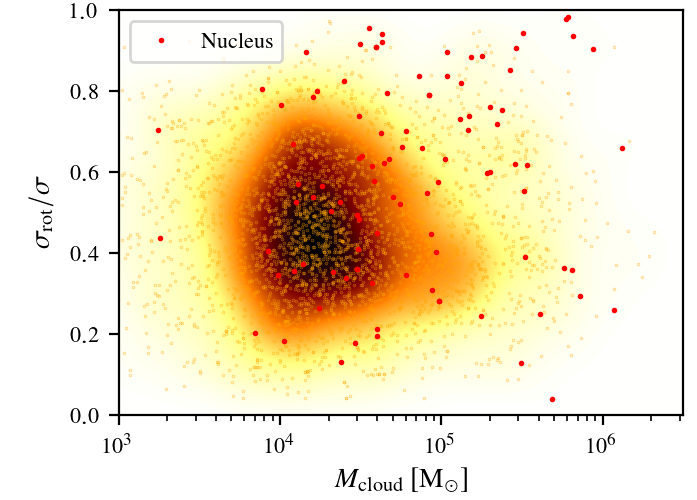}
    \caption{Distribution of the ratio between the velocity dispersion coming from rotation ($\sigma_{\rm rot}$) and the total velocity dispersion ($\sigma$) for the GMCs against their masses. A $\sigma_{\rm rot} / \sigma$ value of $1$ corresponds to clouds whose only contribution to the velocity dispersion is given by rigid body rotation (see Equation \ref{eq:sigmarot}) while the random motions are negligible, while a value approaching $0$ corresponds to clouds that exhibit little rotation compared to their turbulence. The GMCs of the nucleus are shown in red. With the {\em gold-brown color-map} we show the resulting kernel density estimation distribution. }
    \label{fig:RotSigmaDistribution}
\end{figure}

\begin{figure}
	\includegraphics[width=0.94\columnwidth]{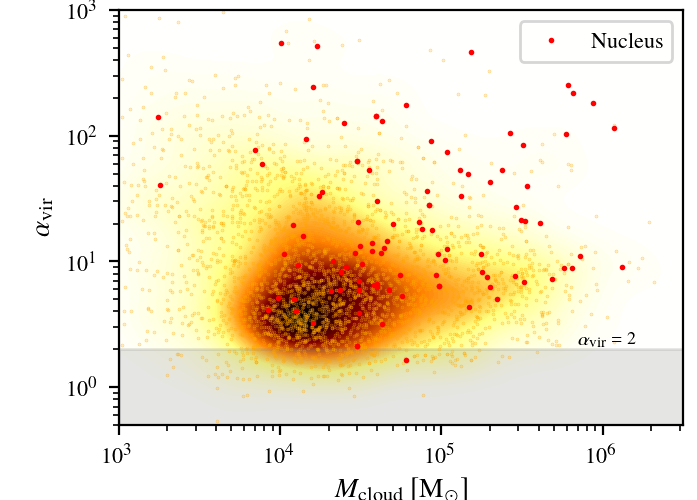}
    \caption{Mass-virial parameter distribution of the clouds in the interacting galaxy. In the {\em gray-shaded region} clouds are considered to be gravitationally bound and collapsing by a simple virial analysis, while clouds above the critical value of $\alpha_{\rm vir} = 2$ are normally considered to be unbound. The GMCs of the nucleus are shown in red. With the {\em gold-brown color-map} we show the resulting kernel density estimation distribution.}
    \label{fig:MAlphaVirScalingRelation}
\end{figure}

We calculate the velocity dispersion 
\begin{equation}
    \mathbf{\sigma} = \left(\frac{\sum_i (\mathbf{v}_i - \mathbf{v}_{\rm com})^2 m_i }{{\sum_i m_i}}\right)^{1/2},
    \label{eq:sigma}
\end{equation}
where $\mathbf{v_i}$ and $m_i$ are the velocities and masses of the Voronoi cells of the cloud and $\mathbf{v}_{\rm com}$ is the velocity of its centre of mass. The sum is extended over all cells within a cloud. We then derive the 1D velocity dispersion:
\begin{equation}
    \sigma_{\rm 1D} = \left(\frac{\sigma_x^2 + \sigma_y^2 + \sigma_z^2}{3}\right)^{1/2}.
\end{equation}
This is closer to what is accessible with observations where we can only measure the velocity dispersion along the line-of-sight. We show the velocity dispersion distribution of the cloud population in Figure~\ref{fig:AlphaVirDistribution}. We see here a bimodal distribution of the velocity dispersion; the secondary peak is a set of clouds with very high velocity dispersion. This is associated with the pathological clouds that are produced by the simulation, which are long-lived clouds that SNe can not disrupt due to the lack of early feedback \citep[see Section \ref{sec:missingPhysics} here and section 5 of][]{Tress+2019}. These objects are long lived and can grow considerably in mass since the feedback cannot halt the collapse, therefore generating massive stellar clusters. They are instead disrupted by cloud collisions eventually. Similar objects appear quite commonly in analogous galaxy scale ISM simulations \citep[e.g.][]{Tasker&Tan2009, Li+2018, Armillotta+2019}.

This calculation includes all motions of the gas in the cloud, including both turbulence and rotation. To see how important rotation is in comparison to random motions, we estimate a rotational velocity dispersion $\mathbf{\sigma}_{\rm rot}$ by computing the angular momentum $\mathbf{L}$ of the clouds (see Section \ref{sec:rotation}) and finding the velocity dispersion of an analogous mass distribution that would rotate as a solid body having the inferred angular momentum. Specifically
\begin{equation}
    \mathbf{\sigma}_{\rm rot} = \left(\frac{\sum_i m_i \mathbf{v}_{{\rm rot}, i}^2}{\sum_i m_i}\right)^{1/2},
    \label{eq:sigmarot}
\end{equation}
where $\mathbf{v}_{{\rm rot}, i} = \mathbf{R}_i \times \mathbf{\Omega}$ is the solid body velocity of the cell. $\mathbf{R}_i$ is the position vector of the Voronoi cells of the cloud with respect to the centre of mass and $\mathbf{\Omega} = \mathbb{I}^{-1} \mathbf{L}$ is the angular velocity of the cloud with inertial tensor $\mathbb{I}$.

We show the ratio between the rotational velocity dispersion and the total velocity dispersion $\sigma_{\rm rot} / \sigma_{\rm 1D}$ in Figure~\ref{fig:RotSigmaDistribution}. We will discuss the rotation of our cloud sample in detail in Section \ref{sec:rotation}.

\subsection{Virial parameter}
\label{sec:virialParam}

\begin{figure*}
    \begin{subfigure}{\textwidth}
	    \includegraphics{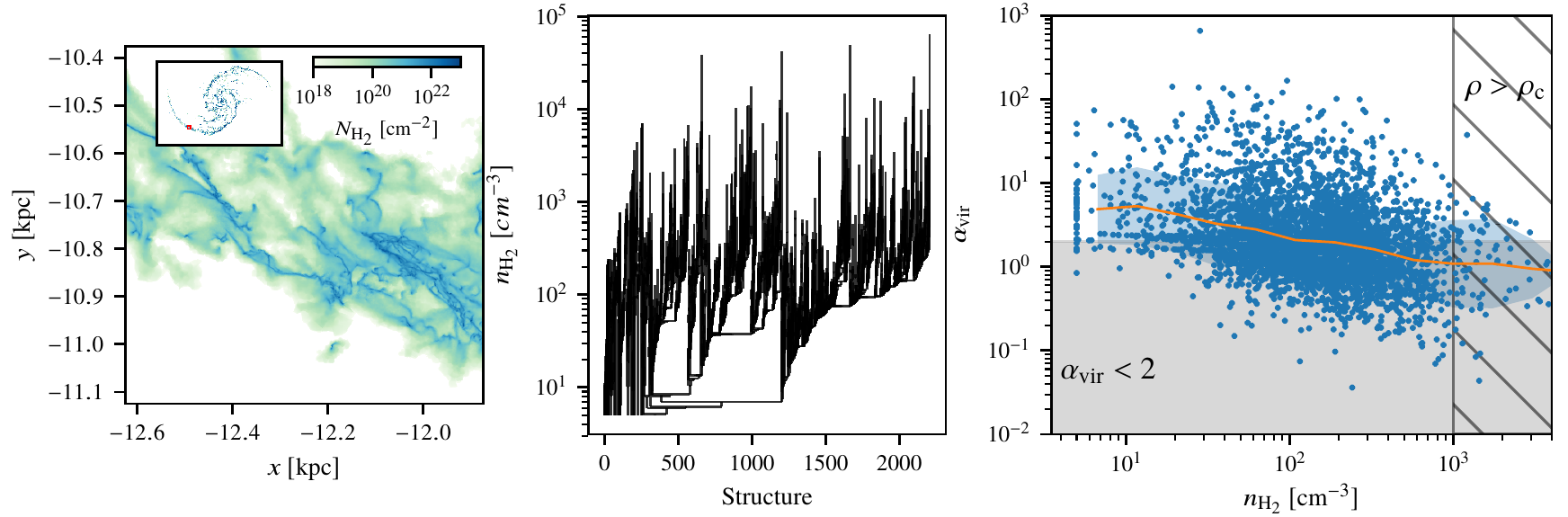}
	\end{subfigure}
    \begin{subfigure}{\textwidth}
	    \includegraphics{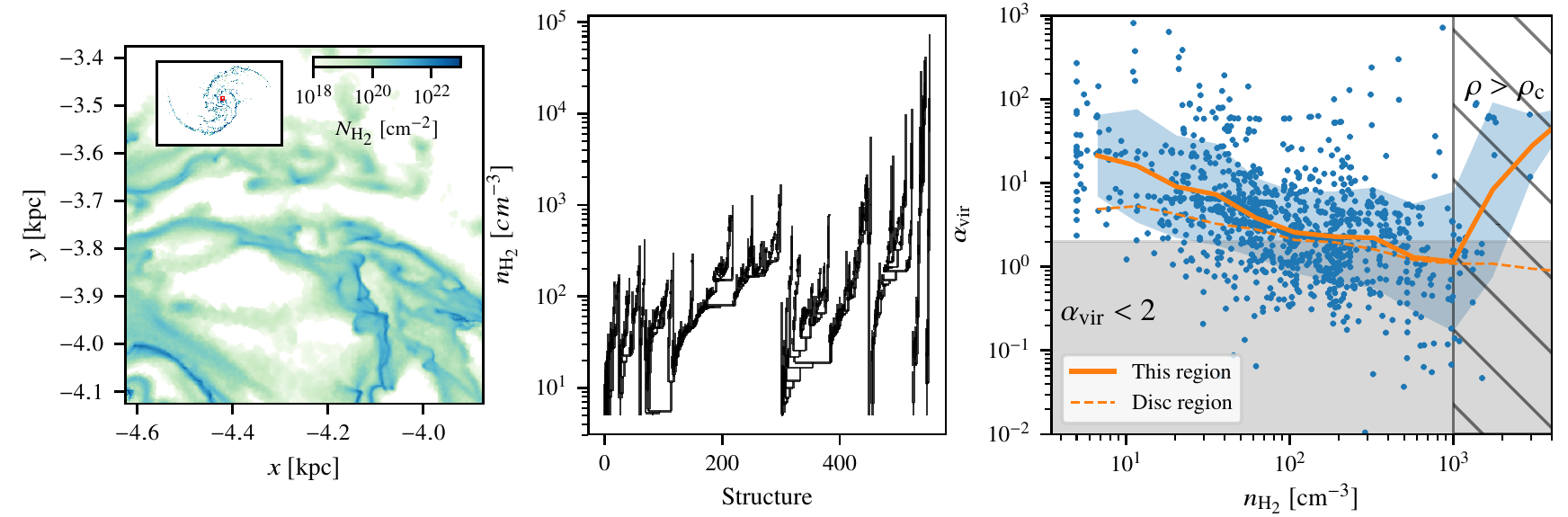}
          \end{subfigure}
          \caption{The virial parameter of structures defined by various molecular iso-density surfaces as a function of molecular density for two different regions in the simulated interacting galaxy, one far out in a spiral arm ({\em top}) and one close to the nucleus ({\em bottom}). We derive the position-space dendrogram ({\em middle panel}) of the region shown in the left panel; for each structure of the dendrogram, we show its virial parameter as a point in the right panel at the density threshold of the structure. The orange line is the binned average of the data and the blue band the $\pm 1\sigma$ deviation from that. For comparison, in the bottom panel we also show ({\em dashed line}) the running average of the region shown in the top panel. The grey band defines collapsing structures based on a virial analysis, while the hatched region shows where the density exceeds the threshold for sink particle creation.}
    \label{fig:WalkTheDendro_AlphaVir}
\end{figure*}

\begin{figure}
	\includegraphics[width=\columnwidth]{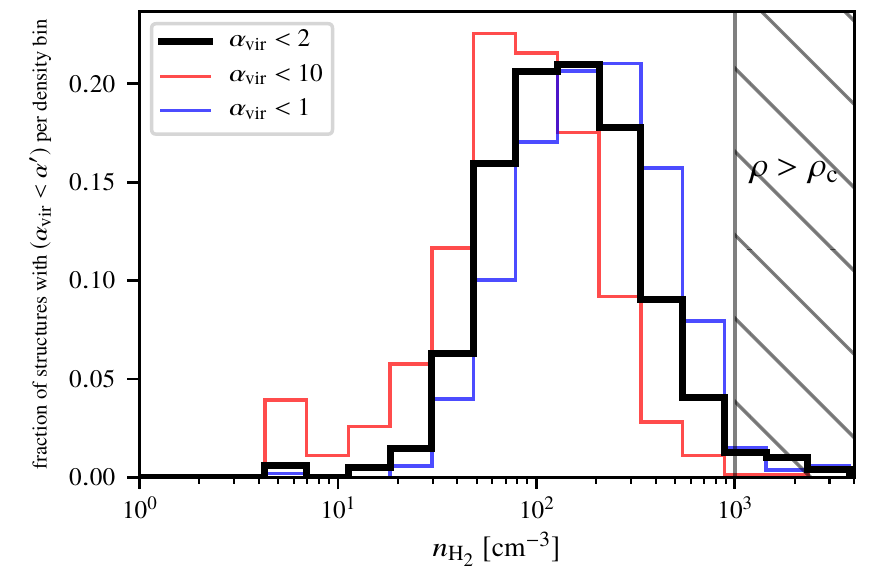}
    \caption{Here we walked the dendrogram of a given region and analysed the lowest structures in the hierarchy of the dendrogram (i.e. least dense) with $\alpha_{\rm vir} < \alpha^\prime$, only containing structures which fulfill the same criterion (see Figure~\ref{fig:CollapsingStructures} to visualise how these structures look like). In this graph we show the density distribution for such structures. The black line is the distribution for such structures with $\alpha_{\rm vir} < 2$ which is commonly accepted to denote gravitationally bound and collapsing regions.}
    \label{fig:AlphaVirOfStruct}
\end{figure}

\begin{figure}
	\includegraphics[width=\columnwidth]{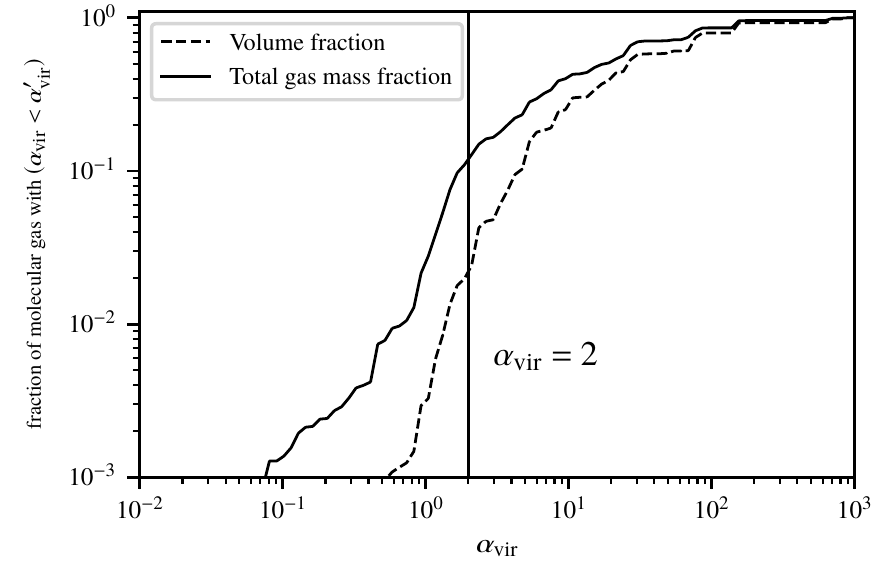}
    \caption{Volume and total gas mass fractions of structures lowest in the hierarchy of the dendrogram (i.e. least dense) having and containing only structures with a virial parameter lower than a given value. These fractions are computed against the total volume/gas mass of the molecular gas, i.e. the gas with $n_{\rm H_2} > 1$~cm$^{-3}$ which is the ISM considered for the dendrogram construction. The vertical line emphasises $\alpha_{\rm vir} = 2$, indicating structures which are generally collapsing. About $10$~\% of the mass and a few percent of the volume is occupied by molecular gas in this condition.  }
    \label{fig:AlphaVirGasFrac}
\end{figure}

\begin{figure}
	\includegraphics[width=\columnwidth]{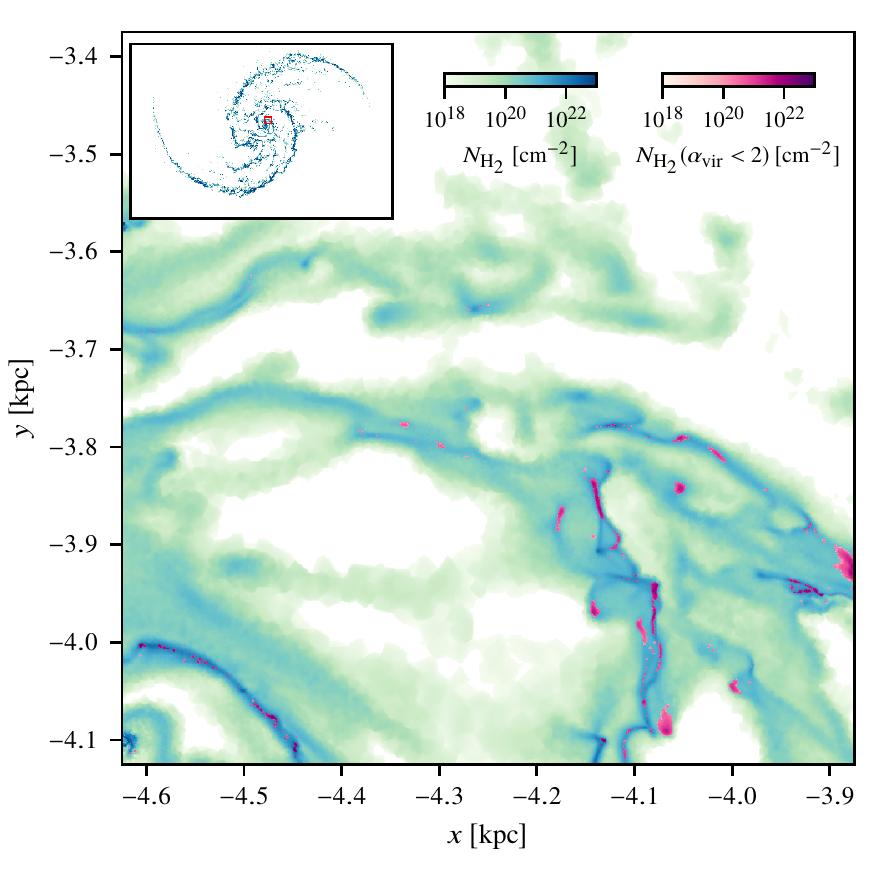}
    \caption{H$_2$ column densities of a given region in the interacting galaxy. The location of the region is shown in the top-left insert. We constructed and analysed the dendrogram of this region and highlight with a red colormap all the structures lowest in the hierarchy of the dendrogram (i.e. least dense) having and containing only structures with $\alpha_{\rm vir} < 2$.}
    \label{fig:CollapsingStructures}
\end{figure}

The virial parameter of a cloud is defined as
\begin{equation}
    \alpha_{\rm vir} = \frac{5 \sigma_{\rm 1D}^2 R_{\rm cloud}}{G M},
    \label{eq:alphavir}
\end{equation}
\citep[see][]{Bertoldi&McKee1992}. This parameter is used as an indication of whether a cloud is collapsing or dissolving. In particular $\alpha_{\rm vir} \propto E_{\rm kin}/E_{\rm pot}$, the ratio of kinetic to potential energy. It can be shown that a Bonnort-Ebert sphere has $\alpha_{\rm vir} = 2.06$, therefore this value is the critical value for stability of non-magnetised clouds and clouds with $\alpha_{\rm vir} \lesssim 2$ are considered to be collapsing. There are issues with this definition, for instance, that clouds in free fall would generally develop velocities from the collapse that raise their $E_{\rm kin}$ and bring the value of $\alpha_{\rm vir}$ closer to unity, making the cloud appear to be stable even though it clearly is not \citep{Ibanez+2016,BallesterosParedes+2018}. Moreover the definition assumes spherical symmetry for the clouds which is clearly an oversimplification (see Figure~\ref{fig:CloudExample}). A stability analysis of GMCs can therefore not solely rely on this parameter. The mass used here $M = M_{\rm cloud} + M_{\rm sink\, gas}$ (see Table~\ref{tab:cloudProperties_clouds}) includes the mass in sink particles within the GMC as well, as they contribute to the local gravitational energy and can influence the stability of the region.

The virial parameter distribution of the cloud populations in the different regions is shown in the right panel of Figure~\ref{fig:AlphaVirDistribution} while the dependence of $\alpha_{\rm vir}$ with the clouds mass is shown in Figure~\ref{fig:MAlphaVirScalingRelation}. Compared to observed structures \citep{Kauffmann+2013}, the simulated GMCs exhibit relatively high virial parameters for their masses, indicating that most of the molecular gas here is gravitationally unbound. Comparable galaxy scale ISM simulations tend likewise to produce predominantly unbound structures \citep[e.g.][]{Dobbs+2011}. 

How can we explain this apparent disagreement with observation? As explained in Section \ref{sec:introduction}, the definition of a GMC is relatively arbitrary and tends to pick out just a few isodensity levels in the hierarchical structure of the molecular ISM. To get a clearer picture of the dynamical state of the entirety of the cold phase, we analyse the virial parameter of all the dendrogram structures of a particular region. In Figure~\ref{fig:WalkTheDendro_AlphaVir} we show $\alpha_{\rm vir}$ of each structure as a function of its threshold density in the dendrogram. In this way we do not favour a specific iso-density surface and can investigate at what typical density the structures transition from a sub- to super-critical state.

We show the results for two regions of the interacting galaxy: the binned average of $\alpha_{\rm vir}$ decreases as the density increases. This is, of course, unsurprising, since it is expected that collapse occurs more easily in higher density regions. The density where the average value of the virial parameter falls below the critical line is $n_{\rm H_2} \sim 10^2$~cm$^{-3}$. We have to highlight, however, that the scatter of $\alpha_{\rm vir}$ is considerable and it is possible to find highly unbound structures even at higher densities. 

At densities exceeding the sink particle formation density threshold, gas only survives if it is highly gravitationally unbound or did not have enough time to be swallowed by a sink particle. This explains why the data points in this area are scarcer and seem to break from the general decreasing trend. It is, however, still interesting to see that even at those densities highly sub-critical structures exist.

One caveat is that we did this analysis for only a few regions. Visual inspection of other such regions, however, suggests similar behaviour. It remains an interesting exercise to study systematically the behaviour of $\alpha_{\rm vir}$ as a function of galactic environment.

The bottom panel of Figure~\ref{fig:WalkTheDendro_AlphaVir} shows the virial parameter density dependence for a region close to the galactic centre. Comparing the trend of $\alpha_{\rm vir}$ to the region farther out in the galaxy disc (top panel), we notice that generally the distribution is shallower and shifted to higher values. This would be expected for a more turbulent and shear dominated region. We leave a more thorough systematic investigation of this statement for future work.

Having a clearer view now of $\alpha_{\rm vir}$ of the molecular gas in all density regimes, we can see that the picture of GMCs being objects in virial equilibrium is rather simplistic; the real ISM might exhibit a more complex structure and variety in internal energies. So why is it that clouds are observationally often found to be close to virialised? One reason is because even collapsing clouds develop velocities that make them look like they are virialised; another reason is survival bias:  clouds with too extreme virial parameters are short-lived. A third reason is selection bias: since we only observe regions where CO becomes bright, we miss the envelopes of clouds, which lower the clouds' virial parameters. This last point explains the difference of the simulated clouds to observed populations, as we select GMCs using the actual H$_2$ density and therefore include even the CO dark gravitationally unbound envelopes. A similar conclusion was reached by \citet{Duarte-Cabral&Dobbs2016} where clouds from the simulation of \citet{Doobs2015} were analysed in H$_2$ and CO, finding that CO traces only the more gravitationally bound parts of clouds. At densities around $n_{\rm H_2} \simeq 10^2$~cm$^{-3}$ where molecular hydrogen becomes CO bright \citep[][]{Clark+2019}, we agree with the observations in finding that the average structure has $\alpha_{\rm vir} \simeq 2$ (see Figure~\ref{fig:AlphaVirOfStruct}). But this should not be confused for a distinctive feature of molecular clouds, but rather a coincidence among the large range of $\alpha_{\rm vir}$ among structures selected at different densities \citep[][]{Beaumont+2013}.

Relative to the total molecular gas, the mass and volume fraction of super-critical structures is comparatively low (see Figure~\ref{fig:AlphaVirGasFrac} and Figure~\ref{fig:CollapsingStructures}). Given the physical conditions simulated here, we find therefore that most structures are gravitationally unbound and only a small percentage of the molecular gas is bound. For the region displayed in the top panel of Figure~\ref{fig:WalkTheDendro_AlphaVir} the volume(mass) fraction of bound gas is 0.033(0.15) while for the region in the bottom panel it is 0.012(0.061). Since only gravitationally bound and collapsing structures could lead to star formation, the low fractions of super-critical molecular gas imply a necessarily low SFE. This is observed in the simulation (see Figure~21 of \citet{Tress+2019}, where depletion times of the molecular gas are $\sim 5\times10^8$~yr) as well as in galaxy observations in general. This suggests therefore that the low galactic SFE is set at the cloud scale.

\begin{figure}
	\includegraphics[width=\columnwidth]{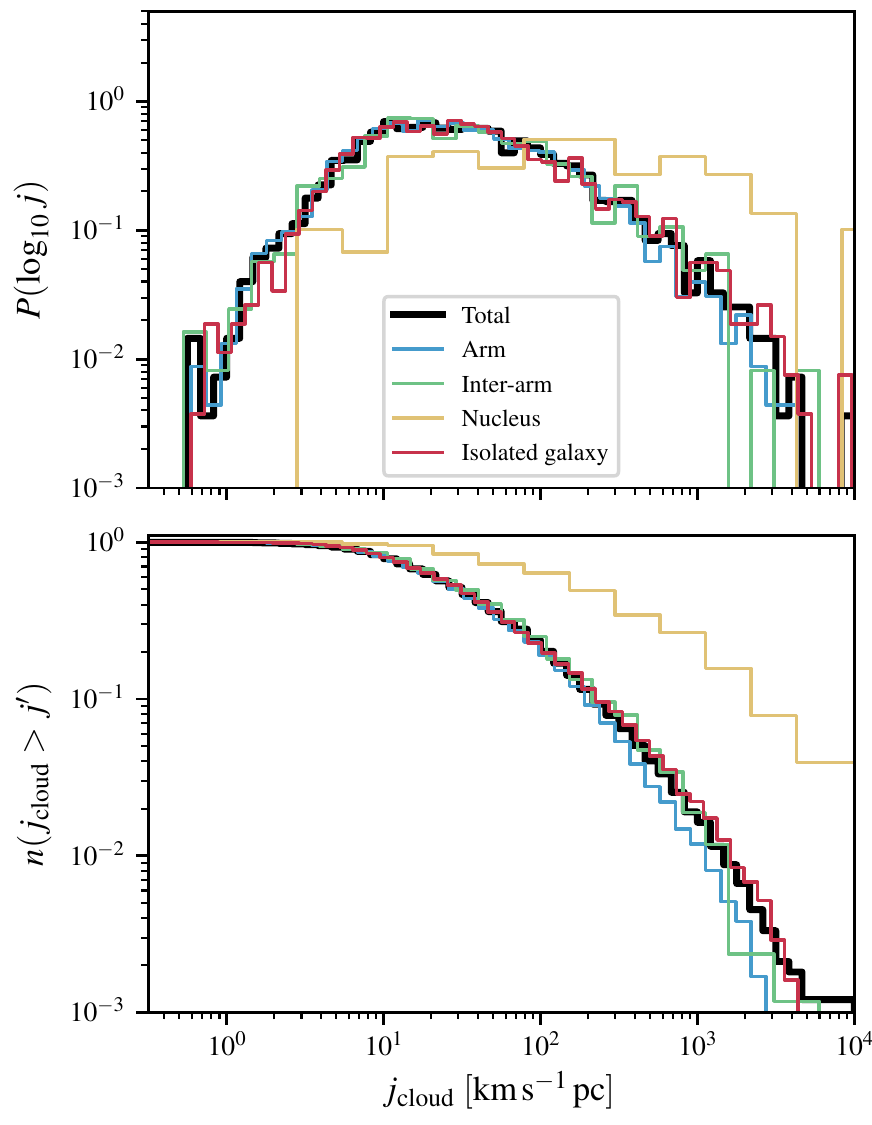}
    \caption{Distribution of specific angular momenta $j_{\rm cloud}$ of the simulated GMC catalog. The different cloud populations are depicted in different colours, consistent with our other figures. The cumulative distribution is shown in the bottom panel.}
    \label{fig:RotationDistribution}
\end{figure}

\begin{figure}
	\includegraphics[width=\columnwidth]{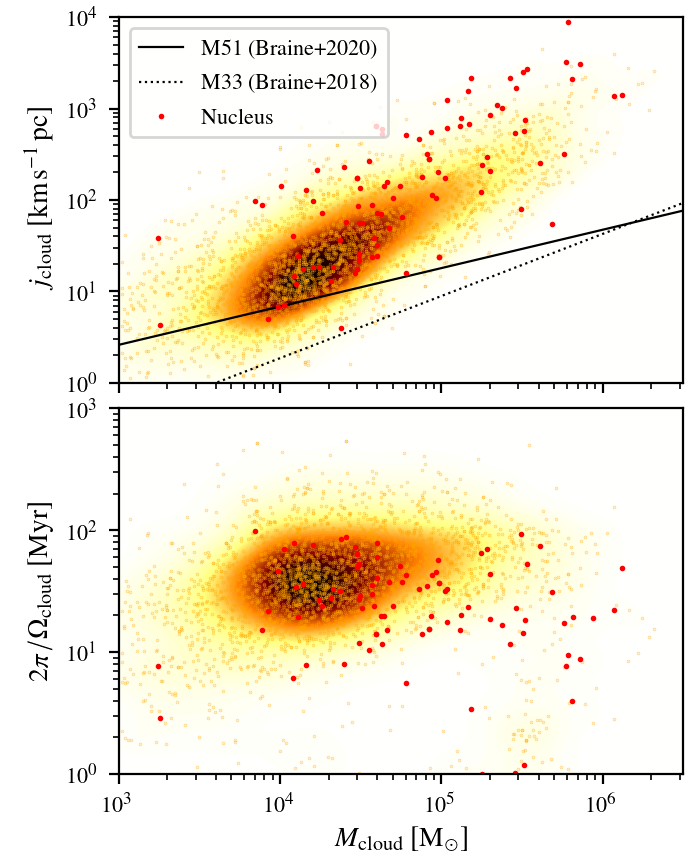}
    \caption{{\em top panel}: Mass-specific angular momentum distribution of the clouds in the interacting galaxy. We show the fits to the observations of M51 and M33 galaxies for comparison. {\em bottom panel}: mass-rotation period distribution of the GMCs. The clouds of the nucleus are highlighted in {\em red}. With the {\em gold-brown color-map} we show the resulting kernel density estimation distributions. }
    \label{fig:MRotScalingRelation}
\end{figure}

\begin{figure*}
	\includegraphics[width=\textwidth]{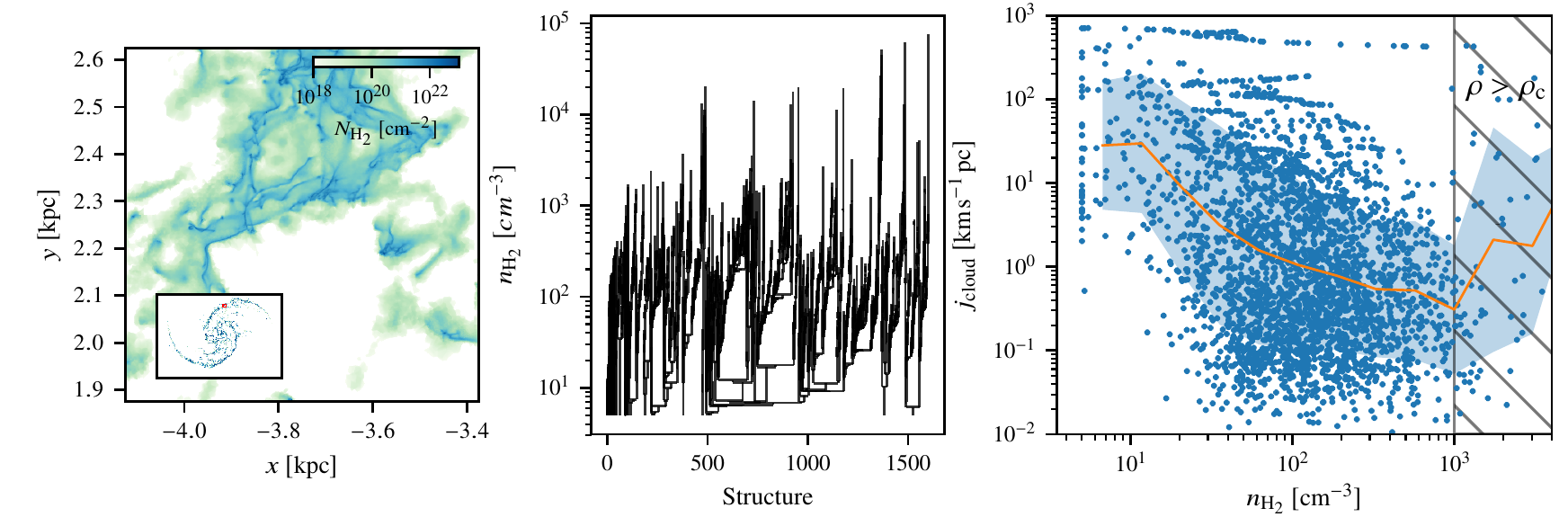}
    \caption{Similar to Figure~\ref{fig:WalkTheDendro_AlphaVir} but here we explore the specific angular momentum of isodensity contours.}
    \label{fig:WalkTheDendro_Rotation}
\end{figure*}

\begin{figure}
	\includegraphics[width=\columnwidth]{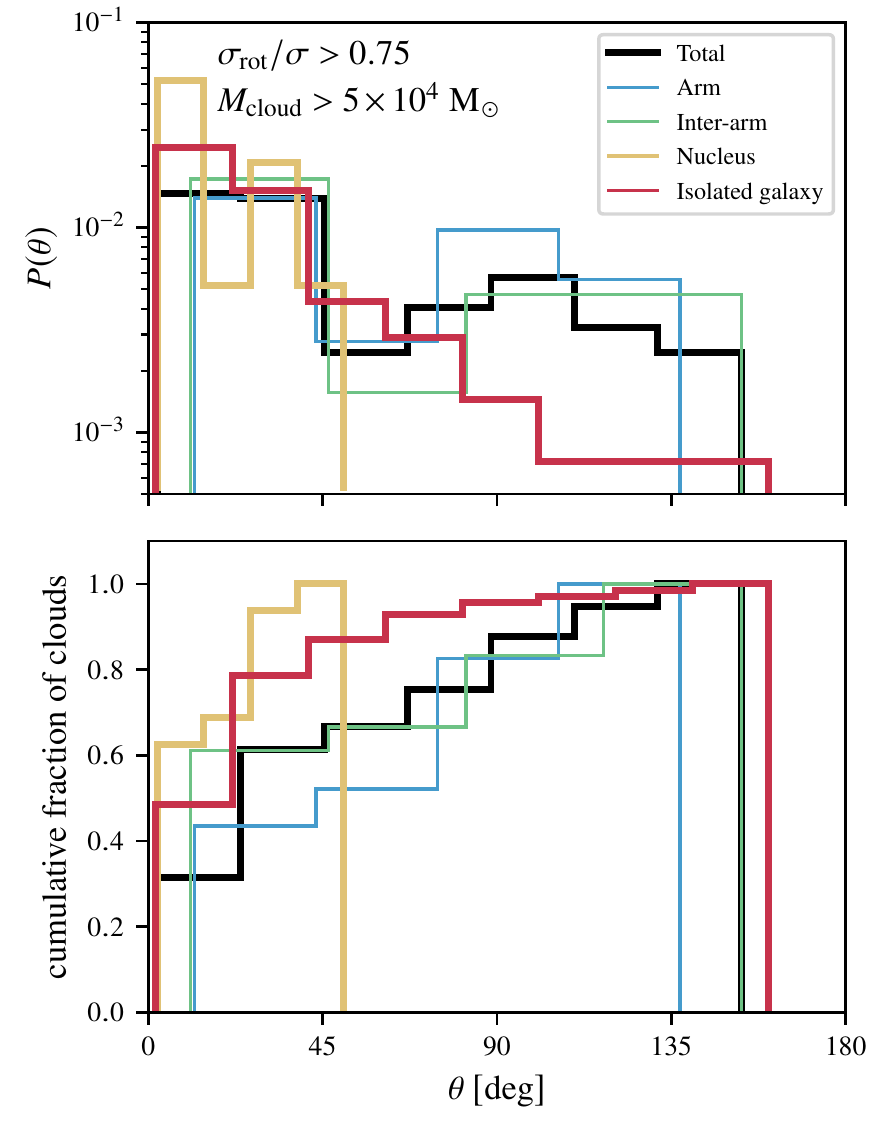}
    \caption{Distribution of the angle between the angular momentum vector of the clouds and the whole galaxy. We only selected rotation dominated clouds with $\sigma_{\rm rot} / \sigma > 0.75$ and excluded low mass ($M_{\rm cloud} < 5\times10^4$~M$_\odot$) clouds for which the local turbulence is likely responsible for their rotation. A value of $0$ denotes clouds completely co-rotating with the galactic disc, while counter-rotating clouds will have $\theta = 180^\circ$. The different cloud populations are depicted in different colours, consistent with other figures. The cumulative distribution is shown in the bottom panel.}
    \label{fig:RotationOfClouds}
\end{figure}

\begin{figure*}
	\includegraphics[width=\textwidth]{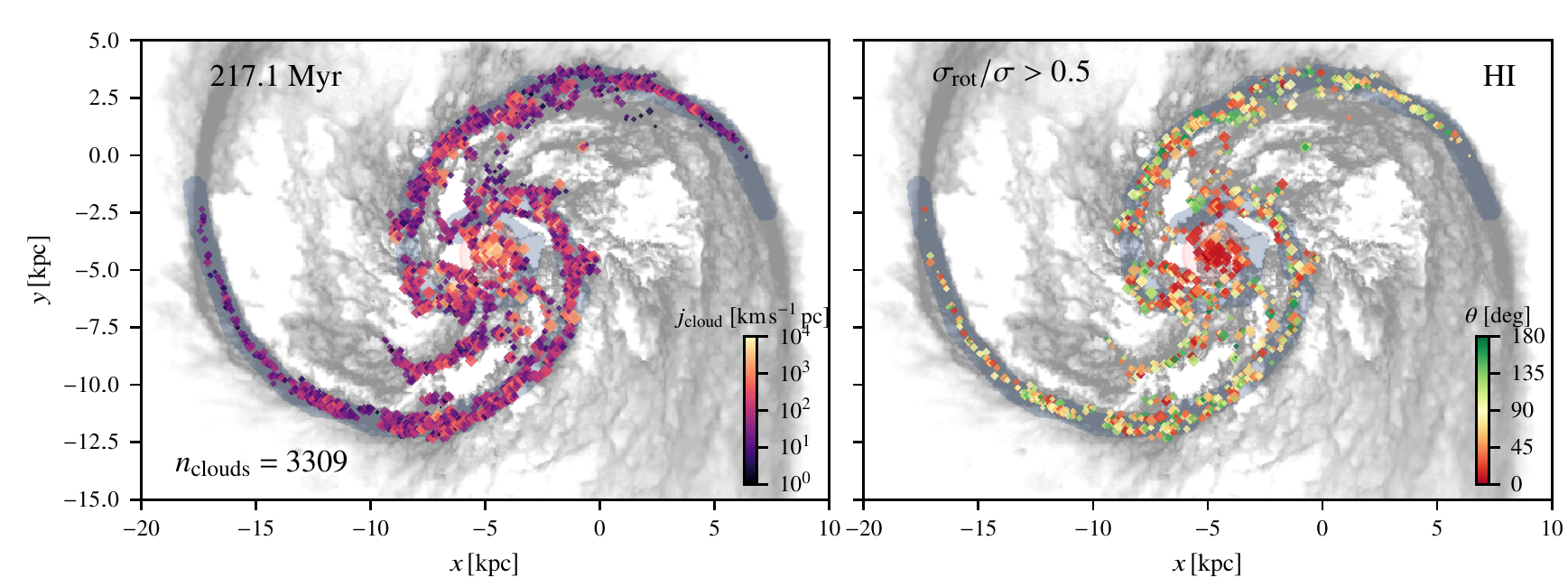}
    \caption{Positions of the GMCs as identified by the cloud-finding algorithm coloured by their specific angular momentum ({\em left hand side panel}) and their inclination with respect to the inclination of the disc ({\em right hand side panel}). Here green clouds are counter-rotating and red clouds are co-rotating with the galaxy. On the right hand side panel we only show clouds with a substantial rotation component, i.e. $\sigma_{\rm rot} / \sigma > 0.5$. }
    \label{fig:CloudRotation}
\end{figure*}

\subsection{Rotation}
\label{sec:rotation}

It has long been suggested that GMCs rotate \citep{Kutner+1977, Blitz1993, Phillips1999}, but it is still unclear whether cloud rotation is dynamically important. Observations suggest that rotational energy is only a small fraction compared to gravitational energy of clouds and so cannot provide any meaningful support against collapse \citep{Rosolowsky+2003, Braine+2018, Braine+2020}. In general, rotational periods seem to always exceed estimated cloud life-times. Environmental variations can however be significant. Clouds in M51 have, for instance, been observed to have three times the specific angular momentum compared to clouds in M33 \citep{Braine+2020}.

The origin of this rotation is also under debate. Most clouds are found to have angular momentum vectors aligned with the plane of the galaxy, supporting the idea that the spin of GMCs is imparted from the galactic rotation curve. Moreover the preferred direction being prograde with the disc rotation, it is believed that it is the orbital rotation that dictates spin direction. In particular in a differentially rotating disc the shear generated by a rising rotation curve will produce prograde clouds by gravitational contraction \citep{Mestel1966}. Of course local turbulence (generated for instance by feedback) has no preferred direction and spinning eddies can be generated regardless of the angular momentum of the disc. The interaction of the ISM with spiral shocks, on the other hand, can establish systematic retrograde vorticity generation \citep{Chernin+1995}. This has been invoked to explain the higher fraction of retrograde clouds in the spiral arms of M51 \citep{Braine+2020}. In the Milky Way and in external galaxies about $30$--$40$\% of clouds are actually counter-rotating such that the simple top-down formation scenario of clouds cannot solely account for cloud formation \citep[see also ][]{Imara&Blitz2011, Imara+2011}.

We compute the specific angular momentum $j_{\rm cloud} = |\mathbf{L}| / M_{\rm cloud}$ of the clouds in our catalogue where $\mathbf{L} = \sum_i m_i \mathbf{r_i} \times \mathbf{v_i}$ is the angular momentum computed over the grid cells contained within the cloud with respect to their centre of mass. We show the distribution of $j_{\rm cloud}$ in Figure~\ref{fig:RotationDistribution}. The clouds in our simulated galaxy have a typical value of $j_{\rm cloud} \simeq 20$~km s$^{-1}$ pc and reach peak values up to $10^4$~km s$^{-1}$ pc. Again there is no evident difference between the clouds of the arm and the inter-arm region, and also the isolated galaxy produces a comparable distribution.

Only the clouds of the central region clearly exhibit a different distribution; here clouds are generally fast rotators and their typical specific angular momentum is more than an order of magnitude greater than disc clouds. Here shearing forces are higher and changes in the galactic rotation velocity curve are significant for scales comparable to the size of a molecular cloud. During gravitational collapse this high shear is then directly translated into rotation of the GMC.

To compare the energies in rotational modes compared to the general velocity dispersion of clouds, in Figure~\ref{fig:RotSigmaDistribution} we show the rotational to total velocity dispersion ratio $\sigma_{\rm rot} / \sigma$ (see Section \ref{sec:sigma} for the definition of $\sigma_{\rm rot}$). On average the rotational velocities constitute about $40$~\% of the total velocity dispersion, but in extreme cases all of the velocity dispersion comes from rotation. This is, to be sure, in part a consequence of the insufficient resolution to properly resolve the turbulent cascade within clouds, which results in an excessive power in large scale rotational modes. On the other hand, though, some of our extreme clouds suffer from inefficient feedback which is unable to disrupt the GMC. The massive sink particles that tend to form in such a situation create long-lived, centrally peaked gravitational fields that are prone to form rotating discs due to dissipation, even though the bulk of our clouds do not suffer from such a problem. 

If GMCs really are the emerging structures of the turbulent cascade, then it is actually not surprising to find that cloud rotation is significant compared to other internal motions. In particular, it is expected from a direct energy cascade that the most power resides in the largest modes. 

We investigate in Figure~\ref{fig:MRotScalingRelation} how the rotation correlates with the mass of the cloud and we see a power-law trend of the specific angular momentum of clouds with increasing masses. The increase in $j_{\rm cloud}$ with mass is consistent with a roughly constant angular velocity of the clouds with mass (bottom panel of Figure~\ref{fig:MRotScalingRelation}). This is suggestive of a top-down formation scenario of GMCs where local shear from the rotation curve plays a major role in driving the rotation as opposed to a bottom-up agglomeration of small clouds in a turbulent medium where instead larger clouds would have a lower chance of having a net rotation.

Observations show positive exponents as well, but compared to M51, our simulation produces a steeper dependence and in general higher values of rotation. We also tried to detach the specific angular momentum     from the definition of a GMC and instead in Figure~\ref{fig:WalkTheDendro_Rotation} we inspected $j_{\rm cloud}$ of iso-density contours as a function of $n_{\rm H_2}$ in a region of the interacting galaxy. There is a general trend of increasing specific angular momentum at lower iso-density levels. This comes plausibly from the large scale shear generated by the rotation curve of the galaxy. A few hierarchical structures in the region shown have high $j_{\rm cloud}$ all the way to the highest density level, i.e. the rotation is dominated by a massive and dense accretion disc-like structure. This can be seen in Figure~\ref{fig:WalkTheDendro_Rotation} from the entries with almost constant $j_{\rm cloud} \sim 10^3$~{km s$^{-1}$ pc} at all density levels.

To study the direction of GMC rotation in our simulations, we selected clouds that had a substantial rotation contribution to their total velocity dispersion, i.e. clouds with $\sigma_{\rm rot} / \sigma > 0.5$, to eliminate the noise of turbulence dominated clouds and slow rotators. Moreover we excluded clouds with low masses ($M_{\rm cloud} < 5\times10^4$~M$_\odot$) for which the local turbulence is more important in regulating their rotation than the galactic shear. In Figure~\ref{fig:RotationOfClouds} we show the distribution of the angles between the angular momentum vector of those clouds and the galaxy. We find that the majority of the clouds are co-rotating with the disc, but the distribution is quite flat with a considerable fraction of retrograde and perpendicular clouds. It is interesting to notice that the interacting galaxy produces a higher fraction of counter-rotating clouds ($\sim 30$~\%) compared to the isolated case ($\sim 10$~\%). \citet{Braine+2020} found a higher retrograde cloud fraction in the arms of the observed M51 galaxy connecting the origin of the counter-rotation to spiral arm passage. Here, however, there is no apparent increase in counter-rotating clouds in the arms (see also Figure~\ref{fig:CloudRotation}). A possible contribution to the increase of counter-rotating clouds is the warp in the disc that the companion galaxy induces as the orbital plane of the two galaxies does not coincide with the plane of the disc. The angular momentum from the encounter could perhaps cascade down to GMC scales and contribute to their rotation direction.

The interaction alters the inclination of GMCs mostly in the outskirts of the galaxy where the forces are greatest, while towards the centre the population stays predominantly co-rotating (see Figure~\ref{fig:CloudRotation}). Moreover, in this region shearing forces are greatest so it is not surprising that here turbulence is unable to produce strongly counter-rotating structures. 

\subsection{Scaling relations}

\begin{figure}
	\includegraphics[width=\columnwidth]{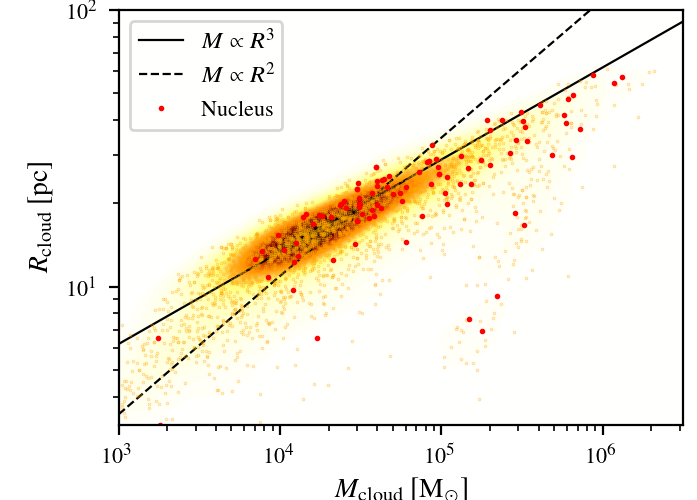}
    \caption{Mass-size distribution of the cloud population for the interacting simulated galaxy. The clouds of the nucleus are highlighted in {\em red}. With the {\em gold-brown color-map} we show the kernel density estimation distribution given the mass and size values of GMCs. The clouds follow a nearly constant density distribution ({\em solid line}) rather than a constant column-density (\em dashed line).}
    \label{fig:RMScalingRelation}
\end{figure}

\begin{figure}
	\includegraphics[width=\columnwidth]{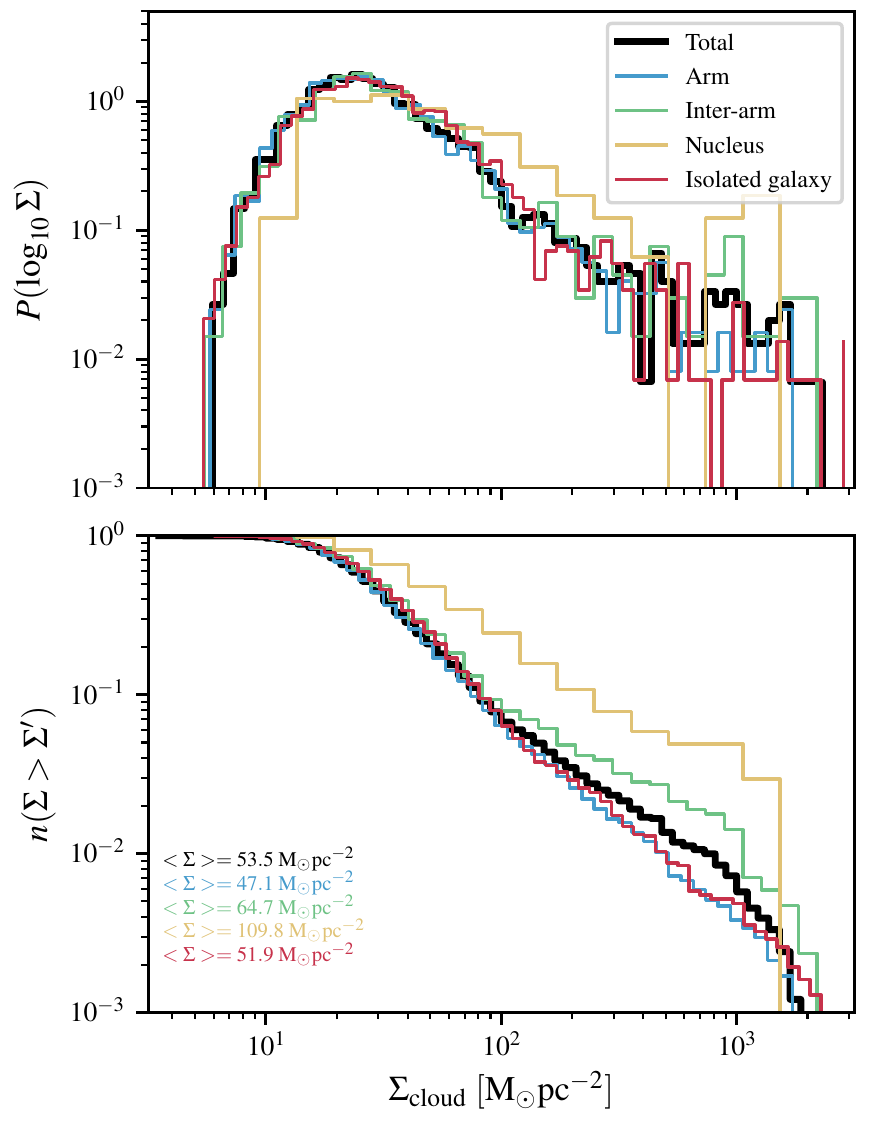}
    \caption{Surface density distribution of the simulated GMC catalog. Here the surface density is derived from the clouds masses $M$ and their size $R$ according to $\Sigma = M/(\pi R^2)$. The different cloud populations are depicted in different colours, consistent with our other figures. The cumulative distribution is shown in the bottom panel.}
    \label{fig:SurfaceDensityDistribution}
\end{figure}

\begin{figure}
	\includegraphics[width=\columnwidth]{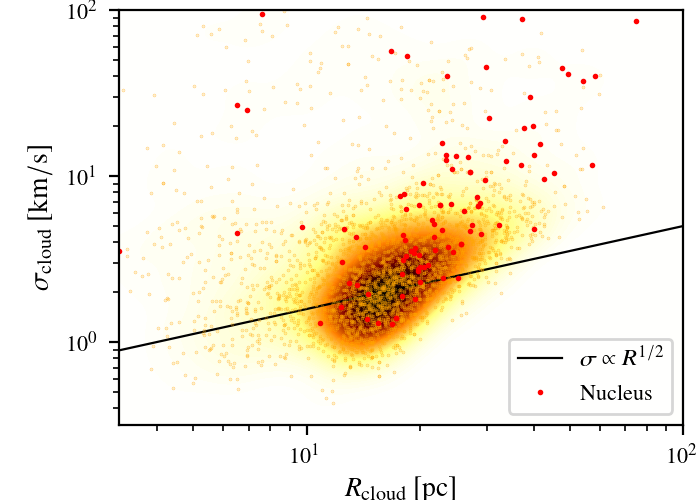}
    \caption{Size-velocity dispersion distribution of the clouds in the interacting galaxy. Symbols and colors are analogous to Figure~\ref{fig:RMScalingRelation}. The {\em solid line} shows $\sigma_{\rm cloud} \propto R_{\rm cloud}^{1/2}$ dependency, typical of virialised structures at constant column density.}
    \label{fig:RSigmaScalingRelation}
\end{figure}

\begin{figure}
	\includegraphics[width=\columnwidth]{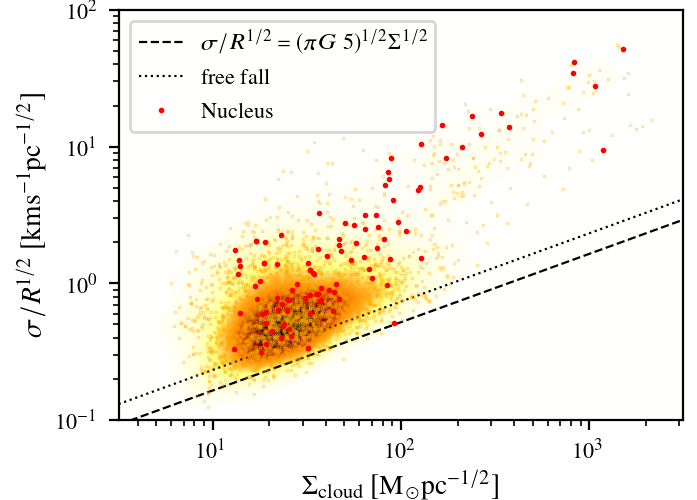}
    \caption{\citet{Heyer+2009} relation for the interacting galaxy cloud population. Symbols and colors are analogous to Figure~\ref{fig:RMScalingRelation}. The {\em dashed line} indicates clouds at virial equilibrium while the dotted line assumes clouds in free fall.} 
    \label{fig:HeyerScalingRelation}
\end{figure}

We analyse here the emerging scaling relations of our simulations. In Figure~\ref{fig:RMScalingRelation} we show the mass-size relation where the mass and the size of the GMCs are defined as described in Section \ref{sec:Masses} and \ref{sec:Sizes}. \citet{Larson1981} first found a relation of the type $M\propto R^2$ that suggested that clouds may have constant mass surface density. This, however, was most likely an observational bias. For our cloud catalogue we find that GMCs span a wide range in surface densities (see Figure~\ref{fig:SurfaceDensityDistribution}) and the relation that we find is rather suggestive of a $M \propto R^3$ type relation, i.e. constant volume density rather than constant surface density. 

We have to stress however that this is likely an artificial result arising from the cloud finding algorithm, which assumes a given volume density threshold to start evaluating the dendrogram as described by \citet{Ballesteros-Paredes+2012}. The average density that the mass-size relation suggests is $\rho\sim7\times10^{-23}$~g~cm$^{-3}$ which corresponds approximately to $n_{\rm H_2}\sim10$~cm$^{-3}$. This is close enough to the threshold density of $n_{\rm H_2,min}=1$~cm$^{-3}$ used by {\sc scimes} such that we cannot rule out a bias from the cloud finding method. 

In Figure~\ref{fig:RSigmaScalingRelation} we show the emerging size-velocity dispersion relation. Clouds in virial equilibrium at constant surface densities would follow a power-law type relation with an exponent of $1/2$. We find a steeper slope for the extracted simulated clouds and large scatter. If the third Larson relation (i.e. constant surface density) does not hold, a dependence of the form $\sigma \propto \Sigma^{1/2}$ is introduced as well in the first Larson relation \citep[see ][]{Heyer+2009}. This is often invoked to explain the large scatter of observed size-linewidth relation of some regions. For the synthetic cloud catalogue we see a great variety in $\Sigma_{\rm cloud}$ and, if we include the surface density dependance, we do approximately retrieve the observed slope for the bulk of our GMCs (Figure~\ref{fig:HeyerScalingRelation}). Other numerical studies of GMCs in a galactic environment come to similar conclusions \citep[see for instance][in particular their Figure 14]{Nickerson+2019}.

Moreover, we saw in Section \ref{sec:virialParam} that the picture of GMCs as virialised objects is rather simplistic and it is therefore misleading to derive scaling relations based on this assumption. The exponents in the size-linewidth power-law relation can vary widely for different targets and, for instance, in M51 no or a weak relation of the linewidth with clouds sizes is observed \citet{Colombo+2014}. A similar power-law relation can also emerge from a turbulent medium, with the slope determined by the inertial cascade \citep{Kritsuk+2013}. 

The structures identified in our simulated galaxies seem mostly gravitationally unbound (see Figure~\ref{fig:AlphaVirDistribution} and \ref{fig:MAlphaVirScalingRelation}) and this is reflected in the bulk of our identified GMCs lying above the Heyer relation (Figure~\ref{fig:HeyerScalingRelation}). Even so, the slope remains close to that inferred for virialised structures. This indicates that gravitationally-driven turbulence is likely substantially contributing to the velocity structure in the clouds, as such motions can mimic virialisation. 

For a subset of clouds, to which also the nuclear GMCs belong, the slope is considerably steeper than the Heyer relation slope. Here other factors are most likely dominant in driving the turbulence in the clouds, such as galactic shear. 

\section{Clouds in the galactic environment}
\label{sec:gal}

\begin{figure}
	\includegraphics[width=\columnwidth]{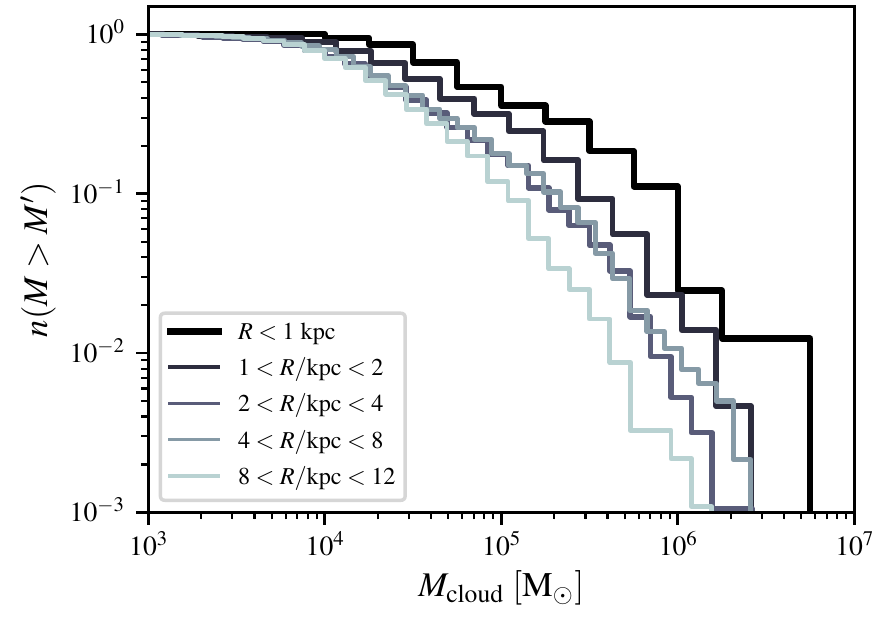}
    \caption{Cumulative mass distribution of the cloud catalogue for different radial bins.}
    \label{fig:MassvR}
\end{figure}

\begin{figure}
    \begin{subfigure}{\columnwidth}
	    \includegraphics[width=\columnwidth]{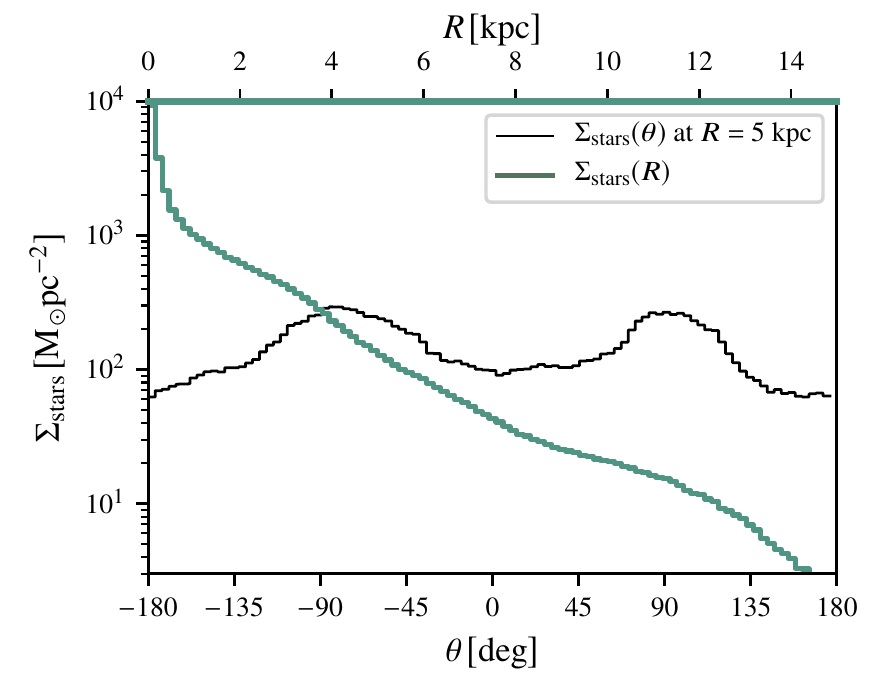}
	\end{subfigure}
	\begin{subfigure}{\columnwidth}
	    \includegraphics[width=\columnwidth]{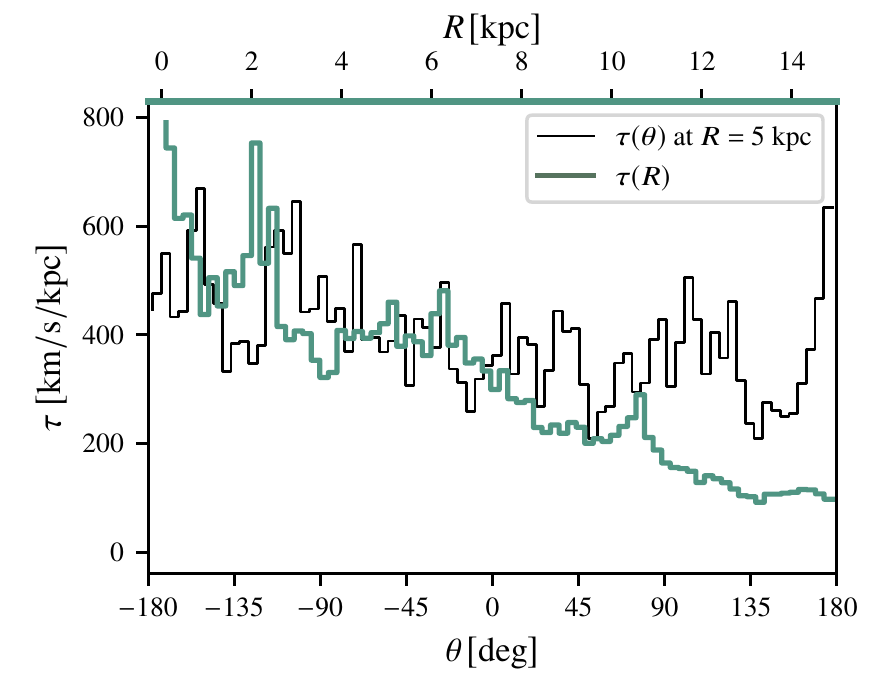}
	\end{subfigure}
    
    \caption{Stellar surface density ({\em top panel}) and mass-weighted average of the shear, as defined in the main text, ({\em bottom panel}) of the interacting galaxy as a function of galactic radius ({\em green line}) and as a function of azimuthal angle at $R=5$~kpc ({\em black line}).}
    \label{fig:sigmaStars}
    
\end{figure}

Spiral arms have always been seen as a major trigger for SF as most molecular gas and SF tracers are observed to be correlated with these galactic scale structures. Correlation does not imply causality however, and it has been shown in some instances that spiral arms can act as a snowplow rather than as a trigger \citep{Tress+2019, Kim&Ostriker2020}. In this sense other galactic parameters such as gas fractions, surface densities and local shear are more important in setting the molecular gas properties. The presence of spiral arms will then just change the distribution of GMCs within the disc without significantly affecting their general properties. This has been evident by the analysis of the general structure of the molecular gas and the SF of these simulations in a previous paper \citep{Tress+2019}, and is emerging from our study of the GMCs in these simulations as well.

The distribution functions of mass, size, velocity dispersion and specific angular momentum of the cloud populations of the arms are virtually indistinguishable from those of the inter-arm clouds. \citet{Duarte-Cabral&Dobbs2016} find that the bulk properties of clouds in their galaxy simulations are also similar for the arm and inter-arm regions, but the tails of some distributions show differences. We also see that, excluding central GMCs, the most massive clouds are associated with the spiral arms (see the tail of the mass distribution in Figure~\ref{fig:MassDistribution}), but are still statistically in agreement with belonging to the same distribution if we consider Poisson noise.  

Moreover the galaxy interaction itself seems to be of little importance in shaping GMCs, as their properties are very similar to those in the galaxy in isolation. \citet{Pettitt+2018} find that for an interacting galaxy comparable to the one presented here, more massive clouds are produced in association with the spiral arms if compared to an isolated galaxy simulation. Their resolution, however, is insufficient to resolve sub-structures in GMCs and here we see that the large GMC structures associated with the arms can be subdivided into a variety of smaller clouds that still follow the same mass function as for the rest of the disc. Moreover \citet{Pettitt+2020} find no change in global cloud properties when performing disc galaxy simulations with different grand design spirals. This would then be an indication that our results are case specific but could be valid to some extent for a different morphology of the spiral arms as well. 

The interaction could however have some importance in determining the rotation direction of clouds, as we observe more retrograde clouds in the interacting simulation. In general some GMC properties could depend more strongly on galactic environment than others, and in particular rotation and aspect ratios are sensitive to the local shear \citep[][]{Jeffreson+2020}. Changes induced by the galaxy interaction are then more evident for those parameters than, for instance, for cloud masses. 

In contrast to our findings, observations of the M51 system show a difference in the GMC population of spiral arms and inter-arm regions \citep{Colombo+2014}. The difference is mainly evident in the masses, where clouds in the arms are generally more massive than inter-arm clouds. The authors attribute this difference to the action of the spiral arms on cloud formation and evolution. We note, however, that the available resolution of the data was considerably lower than what we used to identify clouds. This shows also in the much smaller typical mass of clouds in our simulation compared to the observed ones. The smoothing of the data will blend separate structures into one, which results in more massive objects in a crowded region like a spiral arm. This could explain the disagreement with our results. We plan to perform synthetic observations to compare the extracted clouds more closely with the observations in a follow-up study. 

The central regions of the simulated galaxy, on the other hand, seem to produce clouds that evidently belong to a separate GMC population. Galaxy centres are extreme environments, with high surface densities and extreme shearing forces from the differential rotation. It is therefore not surprising that this is reflected in the evolution of GMCs. We see in Figure~\ref{fig:MassvR} that there is a progressive shift towards more massive clouds for smaller galactic radii. Comparing this to Figure~\ref{fig:MassDistribution} we can see that in our simulations these environmental conditions are not sufficiently different between arm and inter-arm region to affect the formation and evolution of GMCs, but change progressively as we approach more central regions. We can see in Figure~\ref{fig:sigmaStars} that indeed the stellar surface density variation as a function of galactic radius is much higher than for the arm and inter-arm regions. A similar conclusion can be reached by looking at the shear. We define the local shear of the galaxy in two dimensions by using projected quantities. We use the parameter 
\begin{equation}
    \label{tau}
    \tau^2 = \left(\frac{\partial v_x}{\partial y}  +  \frac{\partial v_y}{\partial x} \right)^2 + \left(\frac{\partial v_x}{\partial x} - \frac{\partial v_y}{\partial y} \right)^2,
\end{equation}
which is the magnitude of the eigenvalues of the traceless part of the strain tensor which gives a description of the local shear. Here $v_i$ is the mass weighted mean velocity in the plane of the galaxy. In the bottom panel of Figure~\ref{fig:sigmaStars} we show the mass weighted average of $\tau$ in each radial(angular) bin as a function of galactic radius $R$(azimuthal angle $\theta$). Contrary to the stellar surface density, the variation of the shear as a function of $\theta$ at a given $R$ is comparable to the radial variation. However, there is no strong correlation with the spiral arms as in the inter-arm regions large shear can arise from SN driven bubbles. There is instead a clear trend of increasing shear with decreasing $R$ owing to the galactic rotation. This supports our thesis that the shear plays a greater role in determining the properties of clouds than the morphological presence of a spiral arm. Future work should investigate the correlation of cloud properties not to special places in the galaxy, but to the local physical conditions such as shear, mid-plane pressure and surface densities \citep[see for instance][]{Jeffreson+2018}. 

\section{Caveats}
\label{sec:missingPhysics}

We use this section to discuss the possible implications of the physical ingredients that were not accounted for in the simulations. These are early stellar feedback (such as ionising radiation and stellar winds) and magnetic fields.

SN feedback alone can produce an ISM with reasonable mass and volume fractions in the different thermal phases \citep[][]{Gatto+2015} such that we do not expect the large-scale behaviour of the gas to change significantly if early feedback would be included. As we approach cloud scales and GMC dynamics, however, the effect of early feedback can become substantial \citep[][]{Geen+2015}. We observe this in our simulations with the presence of what we call "pathological" clouds, which are massive, long-lived and often rapidly-rotating dense agglomerations of gas which cannot be disrupted by SN feedback alone. With a proper pre-conditioning of the ISM by early feedback these clouds would evolve differently. But even for more well-behaved clouds in our simulations, the evolution could differ substantially if winds and ionising radiation would have been considered \citep[][]{Rogers&Pittard2013}, in particular for later stages in the lifetime of the cloud. A general trend to be expected is for clouds to have shorter life-times. This could affect clouds during their passage through spiral arms and contribute to a difference in cloud statistics in the arm compared to the interarm region. A massive cloud that in our simulations would survive spiral arm passage and could then be found in the interarm region, with early feedback the same cloud would potentially have been disrupted earlier.

Magnetic fields can also have an effect on the cloud population which may depend significantly on the environment. In general the magnetic field is stronger in the presence of spiral arms \citep[][]{Beck2015, Shanahan+2019, Reissl+2020} and so it could be the driver for inducing a difference in cloud population statistics here. The effect of the magnetic fields varies however for different gas density regimes and the influence is stronger for the diffuse atomic than for the molecular phase \citep[][]{Soler+2020, Soler2021}. We therefore do not expect large differences in the dynamics of the dense gas \citep{Padoan&Nordlund1999, crutcher+2010, bertram+2012} although they will affect certain observational signatures.

We stress that this work is intended as a numerical experiment and not an attempt to faithfully reproduce the natural world. It is rather a useful exercise to learn how the system reacts to certain conditions and physical ingredients. By comparing to real observational data and assessing similarities and differences it gives insight to what elements play fundamental roles in determining observed properties. A simulation which would include all of the physics would have limited scientific advantage as it would obscure the effect of individual physical ingredient on the phenomenon under study. We plan to gradually include further physical processes to explore their effect in a series of future projects. The results presented here will then represent a baseline to compare to.

\section{Conclusions and summary}
\label{sec:summ}

We used the set of simulations presented in \citet{Tress+2019} to study the nature of the cold molecular ISM in the context of an interacting galaxy. These were galaxy scale calculations performed with the hydrodynamic moving-mesh code {\sc arepo}. They include important GMC physics such as a time-dependent chemical network that follows H$_2$ and CO formation and destruction, star formation through sink particles, and SN feedback. They reach sub-parsec resolution in the densest parts of the ISM on scales of an entire galaxy, which is self-consistently evolved throughout the entire time-frame of the interaction. These simulations are therefore particularly useful to study the influence and the effect that galaxy dynamics has on the properties of the molecular phase of the ISM. We focused in particular on the statistical analysis of the emerging GMC population. 

We constructed the dendrogram of the three-dimensional molecular gas distribution. We then used the python package {\sc scimes} to extract molecular clouds at different density levels from a fixed point in time of the simulation. We presented the properties of the structures found in different environments, including their masses, sizes, velocity dispersions, virial parameters, and rotation. 

We can summarise our conclusions as follows:
\begin{itemize}
    \item Despite the interacting galaxy developing prominent spiral arm structures in our model, it does not display the difference in mass function of GMCs of the arm compared to the inter-arm region found in observations. We do, however, see clear differences in molecular cloud properties in the central region of the galaxy, where environmental variables such as shear and surface density have substantially higher values. Our high-resolution maps used to identify clouds enable us to disentangle individual structures in crowded regions such as spiral arms. In contrast, observations at lower resolution and projection effects tend to merge multiple structures, thus introducing bias in the analysis. Our simulations therefore suggest that the structure and dynamics of the molecular ISM is determined by environmental factors such as local shear and mid-plane gravitational forces and surface densities. If the spiral arm cannot significantly alter these conditions, the molecular gas properties remain invariant. 

    \item The cold molecular phase of the ISM is a highly dynamic environment, and GMCs, which are the emerging structures of this phase, reflect this. They exhibit a large range of virial parameters $\alpha_{\rm vir}$, as is expected for a turbulent medium where the energy injection mechanism is not fully coupled to the gravitational energy of the gas. The picture of molecular clouds being virialised objects is therefore misleading and likely the result of observational and selection biases, as a more dynamic and rich picture emerges if we consider the CO dark envelopes of GMCs as well.  We show that, at densities where clouds tend to become CO bright, the average structure shows $\alpha_{\rm vir} \sim 1$, but considering molecular structures at different density levels we can instead find a large spread in $\alpha_{\rm vir}$. Virial analysis shows that only about $10$~\% of the total mass of molecular gas is in a gravitationally bound state that only contains bound structures. The low star formation efficiency of the ISM may well result largely from this low fraction.

    \item We find in our simulations that clouds do not have near constant surface density $\Sigma$, as would be suggested by Larson's scaling relations, but rather span several orders of magnitude in $\Sigma$, similar to the findings of more recent observations that probe larger dynamic ranges \citep[e.g.][]{Hughes+2013, Leroy+2015, Duarte-Cabral+2020}.
    
    \item In our model we find clouds where rotation makes a substantial contribution to their total velocity dispersion. Most of them are prograde with respect to the disc, suggesting that the large scale galactic rotation provides angular momentum at cloud formation through local shear. We find that the interaction with a companion galaxy alters the fraction of prograde clouds, suggesting that some of the orbital angular momentum of the companion cascades down to GMC formation.
\end{itemize}

\section*{Acknowledgements}
The authors are grateful to the anonymous referee whose comments helped to considerably improve the manuscript. The authors acknowledge support from the Deutsche Forschungsgemeinschaft (DFG) via the Collaborative Research Center (SFB 881, Project-ID 138713538) ``The Milky Way System'' (sub-projects A1, B1, B2 and B8) and from the Heidelberg cluster of excellence (EXC 2181 - 390900948) ``STRUCTURES: A unifying approach to emergent phenomena in the physical world, mathematics, and complex data'', funded by the German Excellence Strategy. They also acknowledge funding from the European Research Council in the ERC Synergy Grant ``ECOGAL -- Understanding our Galactic ecosystem: From the disk of the Milky Way to the formation sites of stars and planets'' (project ID 855130). The project made use of computing resources provided by the state of Baden-W\"urttemberg through bwHPC and the German Research Foundation (DFG) through grant INST 35/1134-1 FUGG. Data are stored at SDS@hd supported by the Ministry of Science, Research and the Arts Baden-W\"urttemberg (MWK) and the German Research Foundation (DFG) through grant INST 35/1314-1 FUGG. RJS gratefully acknowledges an STFC Ernest Rutherford fellowship (grant ST/N00485X/1)  and HPC from the Durham DiRAC supercomputing facility (grants ST/P002293/1, ST/R002371/1, ST/S002502/1, and ST/R000832/1). M-MML was partly supported by US NSF grant AST18-15461. ADC acknowledges support from the Royal Society through a University Research Fellowship (URF/R1/191609).
This research made use of {\sc astrodendro}, a Python package to compute dendrograms of Astronomical data (http://www.dendrograms.org/) and of {\sc scimes}, a Python package to find relevant structures into dendrograms of molecular gas emission using the spectral clustering approach.

\section*{data availability}
The data underlying this article will be shared on reasonable request to the corresponding author. The properties of the cloud population in the case of the interacting and isolated galaxy are available as online supplementary material.




\bibliographystyle{mnras}
\bibliography{bibliography}


\appendix

\section{Different times}
\label{sec:differentTimes}

\begin{figure}
	\includegraphics[width=\columnwidth]{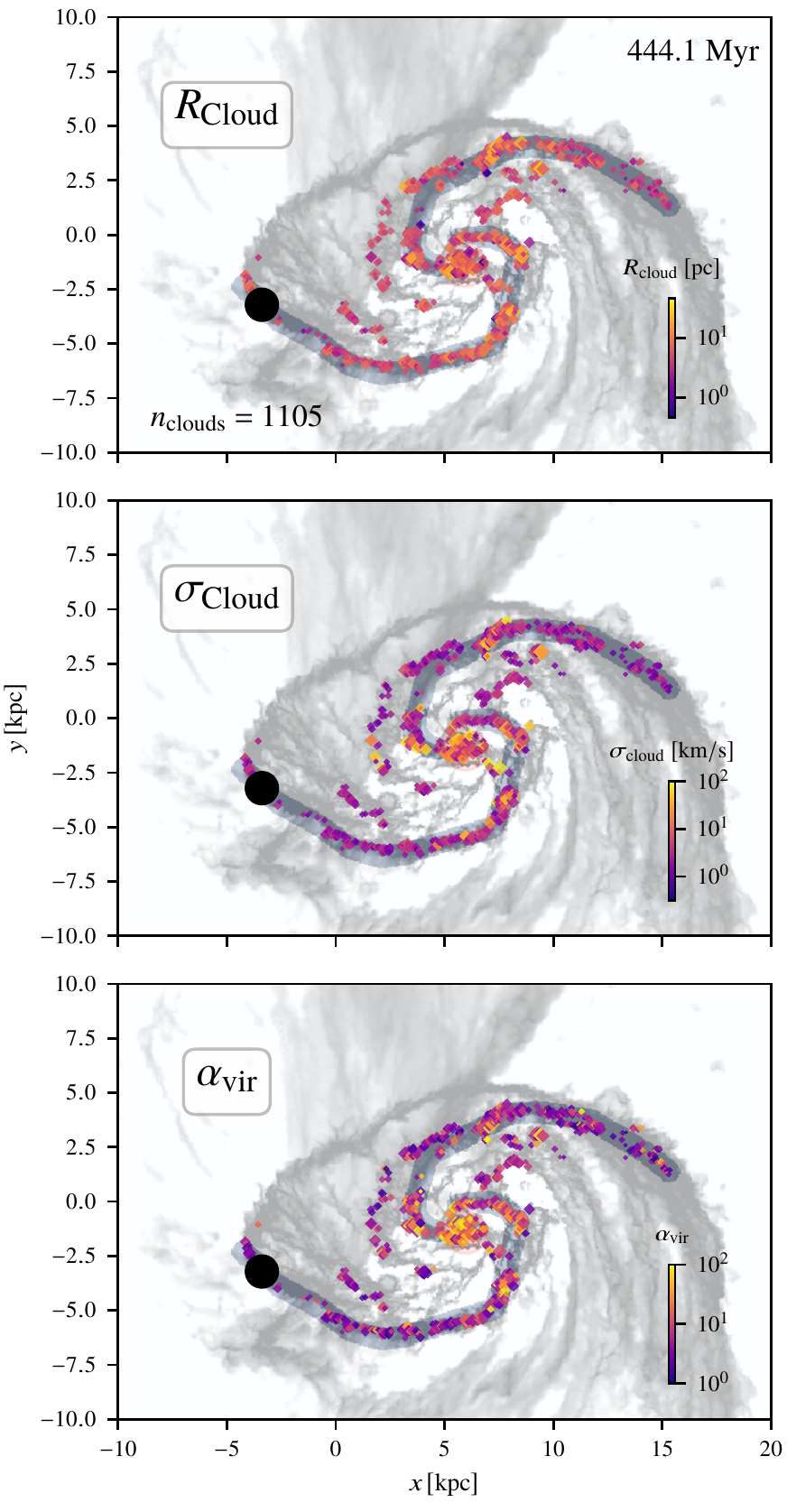}
	\caption{Same as Figure~\ref{fig:cloudPropertiesMap} but for the interacting galaxy simulation at a later time. The position of the companion galaxy, which is modelled as a single massive particle, is shown as the big black filled circle in the three panels. }
    \label{fig:CloudRSigmaAlphaVir_int_2}
\end{figure}

\begin{figure}
	\includegraphics[width=\columnwidth]{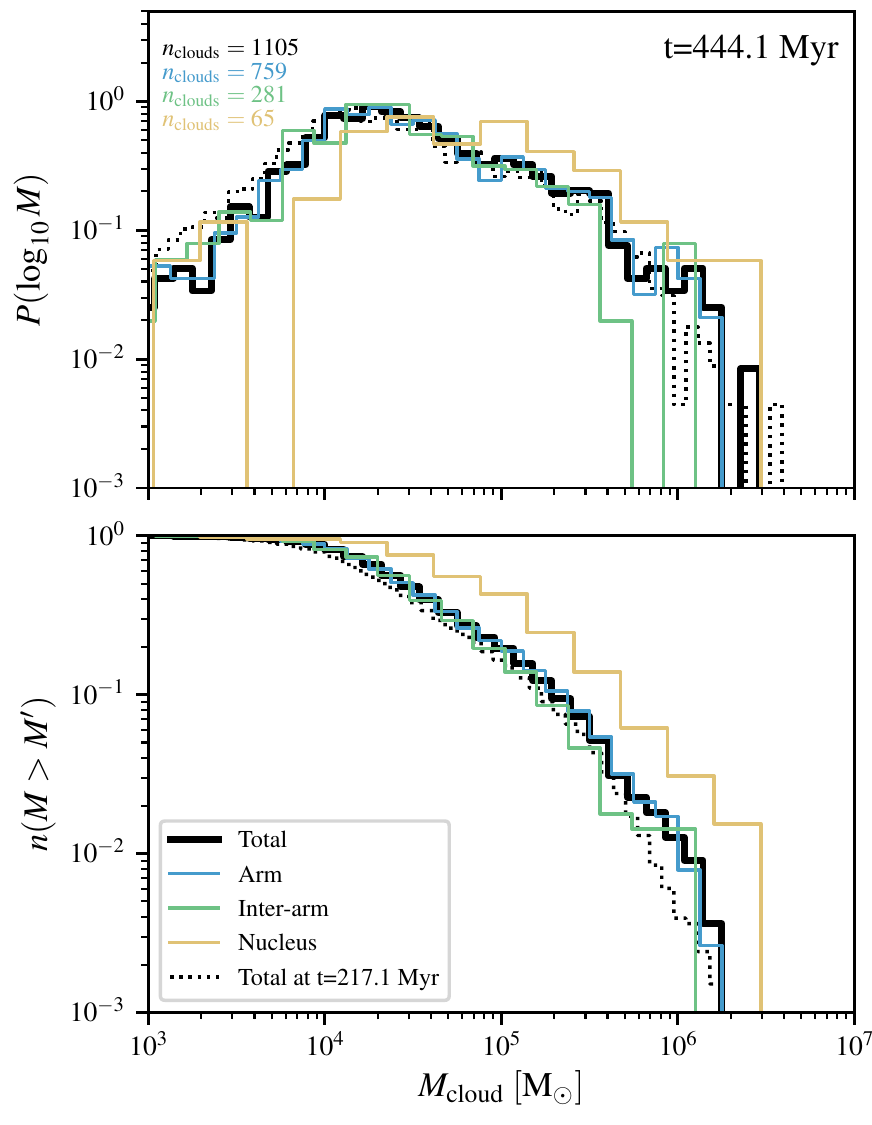}
	\caption{Similar to Figure~\ref{fig:MassDistribution} we show here the mass distribution of the clouds identified in the interacting galaxy at a simulation time of $t=444.1$~Myr. We show the structures attributed to the centre, the arm and the inter-arm region as well as the total mass distribution of the cloud catalogue at this time ({\em solid black line}). For comparison we also show the mass distribution of the clouds identified at an earlier time snapshot, used and described in the main text of this work ({\em dotted black line}). }
    \label{fig:massFunction_2}
\end{figure}

Compiling a cloud catalogue for a given snapshot is computationally expensive enough that we could not afford to extract the clouds for every simulation snapshot. This raises the question, however, whether our results are valid throughout the evolution of the galaxy. It could be argued, for instance, that the spiral arms of the galaxy at the time chosen for the main analysis are dynamically young and still in the process of developing. The gas could then not have had enough time to leave the freshly formed arms which could be the cause for the small difference in cloud properties between arm and inter-arm regions that we observed and described in the main body of this work. This might be especially valid for the outer parts of the disc where the rotation periods are long. To address this we extracted here the clouds of a second snapshot at a later stage of the simulation ($t=444.1$~Myr). The configuration of the simulated galaxy and the companion at this time corresponds to the one of the M51 system which our model was designed to roughly reproduce. We defined the spiral arm, inter-arm region and nucleus in the same way as described in the main text (see Section \ref{sec:cloudCatalogue}). 

We show in Figure~\ref{fig:CloudRSigmaAlphaVir_int_2} the locations of the clouds found coloured by their size, velocity dispersion and virial parameter overlaid to the HI column density of the galaxy. Moreover in Figure~\ref{fig:massFunction_2} we look at the mass distribution of the clouds. A total of $1167$ clouds were found, a considerable smaller amount with respect to the snapshot at $t=217.1$~Myr. This is to be attributed to the gas depletion due to the intense SF as well as a slightly higher number of massive clouds compared to earlier times.

If we compare the solid to the dotted line in Figure~\ref{fig:massFunction_2} we see indeed a small deviation at the high-mass end, suggesting that the interaction scenario might slightly favour more massive clouds as the merger progresses. The difference, however, is relatively small.

In the main text we argued that the spiral arms were acting more as a snow-plow rather than triggering new cloud formation. In this sense they mainly collected the clouds from the inter-arm regions without substantially affecting their properties. The same conclusion can be drawn from this snapshot as the cloud population of the inter-arm is very similar to the arm clouds. Only the central regions seem to be systematically of higher mass.

We also selected a few random patches of the interacting galaxy at different times and identified the clouds there. We compared the key properties of those clouds found to the cloud catalogue of the main snapshot analysed. No major differences were detected in the cloud populations. No major changes were expected as the galactic conditions and the star formation differed only slightly during the time period considered (see Figure~15 of \citet{Tress+2019}). This confirms that our conclusions are not the result of a particular choice of time, but are general for the type of galaxy and environmental conditions.   


\bsp	
\label{lastpage}
\end{document}